\documentclass{pasj01}

\Received{XXXXX}
\Accepted{XXXXX}
\Published{$\langle$publication date$\rangle$}

\begin{document}

\title{Characterization of M dwarfs Using Optical Mid-Resolution Spectra for Exploration of Small Exoplanets}
\author{Yohei \textsc{Koizumi} \altaffilmark{1}, 
	    Masayuki \textsc{Kuzuhara} \altaffilmark{2, 3}, 
	    Masashi \textsc{Omiya} \altaffilmark{2, 3}, 
	    Teruyuki \textsc{Hirano} \altaffilmark{1}, 
	    John \textsc{Wisniewski} \altaffilmark{4}, 
	    Wako \textsc{Aoki} \altaffilmark{3}, 
	    and Bun'ei \textsc{Sato} \altaffilmark{1}}

\altaffiltext{1}{Department of Earth and Planetary Sciences, Tokyo Institute of Technology, 2-12-1 Ookayama, Meguro-ku, Tokyo 152-8551, Japan}
\altaffiltext{2}{Astrobiology Center, NINS, 2-21-1 Osawa, Mitaka, Tokyo 181-8588, Japan}
\altaffiltext{3}{National Astronomical Observatory of Japan, NINS, 2-21-1 Osawa, Mitaka, Tokyo 181-8588, Japan}
\altaffiltext{4}{Homer L. Dodge Department of Physics and Astronomy, University of Oklahoma, 440 W. Brooks Street, Norman, OK 73019, USA}

\email{koizumi.y.ac@m.titech.ac.jp}

\KeyWords{stars: late-type, stars: low-mass, stars: fundamental parameters}

\maketitle

\begin{abstract}
	We present the optical spectra of 338 nearby M dwarfs, and compute their spectral types, effective temperatures ($T_{\mathrm{eff}}$), and radii. 
	Our spectra have been obtained using several optical spectrometers with spectral resolutions that range from 1200 to 10000.
	As many as 97\% of the observed M-type dwarfs have a spectral type of M3--M6, with a typical error of 0.4 sub-type, among which the spectral types M4--M5 are the most common.  
	We infer the $T_{\mathrm{eff}}$ of our sample by fitting our spectra with theoretical spectra from the PHOENIX model.  
	Our inferred $T_{\mathrm{eff}}$ is calibrated with the optical spectra of M dwarfs whose $T_{\mathrm{eff}}$ have been well determined with the calibrations that are supported by previous interferometric observations.
	Our fitting procedures utilize the VO absorption band (7320--7570 \AA) and the optical region (5000--8000 \AA), yielding typical errors of 128 K (VO band) and 85 K (optical region). 
	We also determine the radii of our sample from their spectral energy distributions (SEDs).
	We find most of our sample stars have radii of $<$ 0.6 $R_\odot$, with the average error being 3\%.
	Our catalog enables efficient sample selection for exoplanet surveys around nearby M-type dwarfs.
\end{abstract}

\section{Introduction} \label{sec:intro}
	In recent years, M dwarfs have been gathering attention as attractive targets to detect Earth-mass planets in habitable zones.
	M dwarfs are main-sequence stars with effective temperatures ($T_{\mathrm{eff}}$) ranging from 2300 to 3900 K (e.g., \cite{Rajpurohit2013}) and masses ranging from 0.09 to 0.6 $M_\odot$, approximately (e.g., \cite{Kraus2007}).
	The low temperatures cause the habitable zones around these stars to be located closer to the central stars compared to hotter stars.
	Combining the small orbital distance and the low stellar mass, the stellar radial-velocity variations caused by an Earth-mass planet in the habitable zone has an amplitude comparable to $\sim$1 m/s, which is ten times larger than that of a solar-mass star hosting an Earth-mass planet in the habitable zone (e.g., \cite{Burke2014}).
	Therefore, it is easier to detect Earth-mass planets in the habitable zones around M dwarfs than around Sun-like stars.
	
	Rocky planets are expected to be more common around M dwarfs as compared to Sun-like stars \citep{Dressing2013}, and M dwarfs comprise more than 70\% of stars in the Solar neighborhood \citep{Henry1994,Reid2004}.
	This enhances the probability of detecting habitable planets around M dwarfs in the solar neighborhood.
	For these reasons, several groups are planning planet-search programs targeting M dwarfs using state-of-the-art high-dispersion spectrographs such as IRD \citep{Kotani2018}, SPIRou \citep{Artigau2014}, HPF \citep{Mahadevan2012}, and CARMENES \citep{Quirrenbach2014}.
	
	Accurate characterization of fundamental stellar properties is very important for planet-search programs because the estimation of planetary physical properties such as mass and radius directly depend on these stellar properties.
	These fundamental stellar properties are also crucial when we compare the properties of planetary systems between various types of stars.
	For example, it is known that the occurrence rate of giant planets is correlated to the stellar mass (e.g., \cite{Johnson2010,Gaidos2013}), and that of small planets increases as the stellar temperature decreases \citep{Howard2012}.
	It should be stressed here that stellar properties such as mass, radius, and luminosity are all correlated with the stellar $T_{\mathrm{eff}}$ (e.g., Mann et al. 2013b; 2015).
	Therefore, the precise determination of stellar temperatures is necessary in exoplanet surveys.
	
\begin{figure}
		\centering
		\includegraphics[width=\columnwidth]{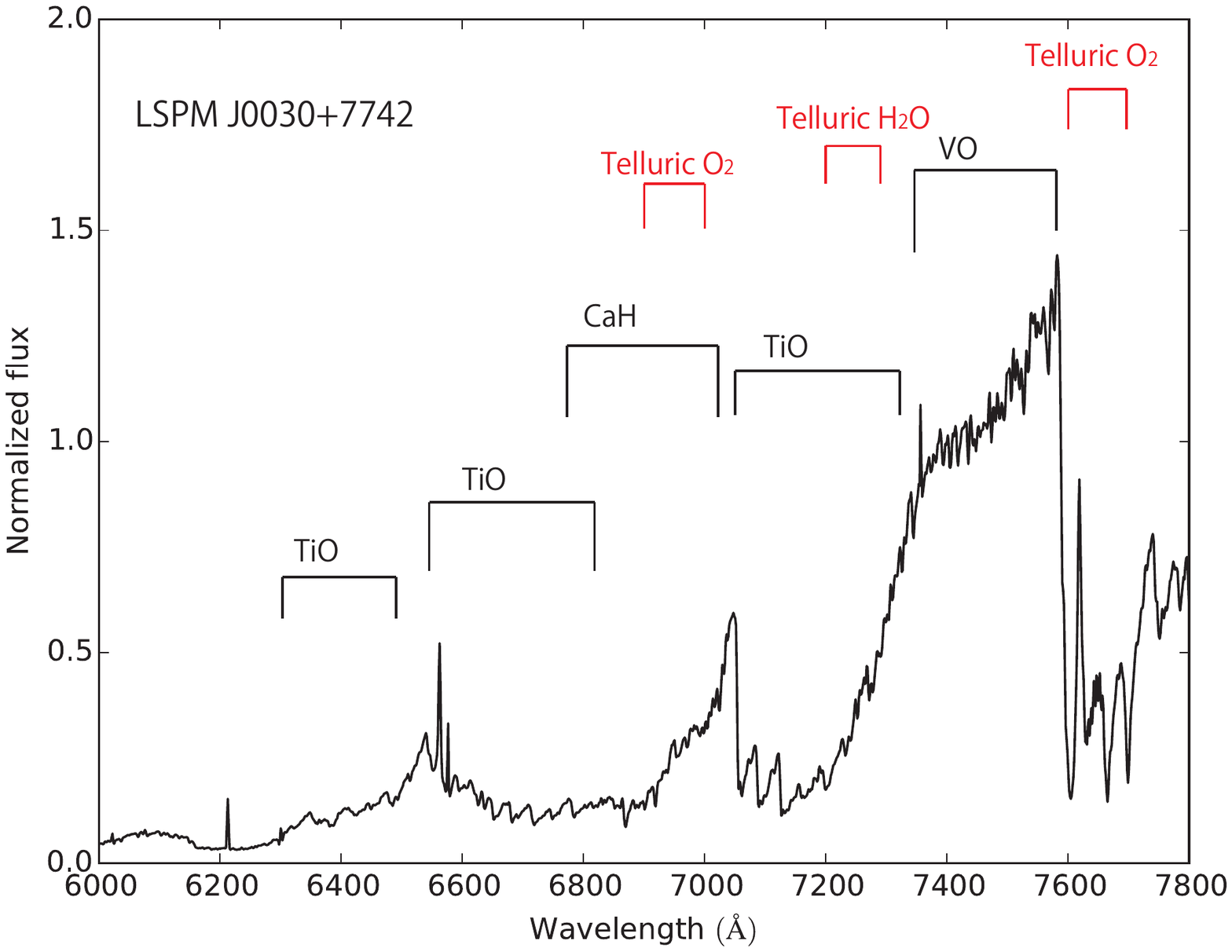}
		\caption{
			Molecular absorption bands in optical spectra of M dwarfs (e.g., \cite{Kirkpatrick1991}).
			The spectrum plotted in this figure was obtained in our WIYN observations (see section \ref{sec:data_obs}).
             		The red bands represent telluric absorptions.
			\label{fig_molecularbands}
		}
	\end{figure}

	It is more challenging, however, to determine the stellar temperatures of M dwarfs than those of solar-type stars.
	For M dwarfs, molecular absorption bands (mainly metal oxides and hydrides such as TiO and CaH) are dominant in their optical and near-infrared spectra because of their low $T_{\mathrm{eff}}$ \citep{Bessell1991,Kirkpatrick1991}.
	Figure \ref{fig_molecularbands} shows typical molecular absorption features in the optical spectra of M dwarfs. 
	For solar-type stars, their physical properties can be determined by using equivalent widths of atomic absorption lines.
	However, the continuum region in M dwarf spectra cannot be accurately determined because of strong and complex absorption bands (see Figure \ref{fig_molecularbands}).
	It is then notable that the synthetic model atmosphere for low mass stars is still incomplete as discussed in \citet{Veyette2017}, making it further difficult to characterize those cool dwarfs.
	Hence, it is crucial to validate how synthetic model atmospheres are applied to observed spectra of M dwarfs given the model uncertainties.
	Actually, the empirically calibrated methods to constrain the properties of M dwarfs were provided by Mann et al. (2013b; 2015).
	They also applied their methods to a lot of M dwarfs by comparing their optical and near-infrared medium-resolution spectra to synthetic model spectra, providing the $T_{\mathrm{eff}}$ over a range of 2600--4200 K with the precision of $\sim60$ K.
	
	Regardless of the improvements on the synthetic model spectra, there are still some wavelength regions that are not suitable for estimation of stellar properties (e.g., \cite{Reyle2011}).
	Furthermore, it is possible that the existing techniques of stellar characterizations are not suited for the specifications of available instruments (e.g., narrow wavelength coverages).
	We thus need to develop the techniques that can be applied to different instrumental specifications, with which various telecopes and instruments can be applied to a variety of M dwarfs such as distant M dwarfs.
	
	The radii of bright and nearby M dwarfs can be determined using their parallaxes and angular diameters measured by interferometric observations, as shown by \citet{Boyajian2012}, who measured the stellar radii of a set of early M dwarfs.
	In the same work, they also determined stellar luminosities by combining bolometric fluxes with parallaxes. Stellar masses have also been determined from the empirical relations between the absolute $K$ magnitudes and masses (e.g., \cite{Henry1993,Delfosse2000}).
	Mann et al. (2013b; 2015) derived the relations between $T_{\mathrm{eff}}$ and stellar properties (radius, mass, and luminosity) by using the measurements of \citet{Boyajian2012}.
	However, M dwarfs are generally small and faint, which makes direct radius measurements using interferometry difficult for most M dwarfs.
	
	M dwarfs' radii are also measured by observations of eclipsing binary stars; however, these radii are 5--15\% larger than that predicted by theory (e.g., \cite{Torres2002,Kraus2011}).
	The lack of precise measurements of parallaxes for M dwarfs was one of the major sources of error when estimating the properties of M dwarfs.
	Because Gaia \citep{Gaia2018} now provides precise parallaxes of nearby M dwarfs, we can improve the estimation of M dwarf radii.
	
	In this paper, we determine the $T_{\mathrm{eff}}$ of M dwarfs by using medium-resolution optical spectra, which are more easily available than their infrared spectra, and determine their radii using photometric data and parallaxes.
	Our method is calibrated with M dwarfs whose $T_{\mathrm{eff}}$ are well determined by previous studies that employed interferometric observations.
	For spectra with limited or narrow wavelength coverages, we find that the $T_{\mathrm{eff}}$ estimation is refined by utilizing only VO absorption features if available.
	Compared with TiO absorption, this absorption can well reproduce temperatures determined in \citet{Mann2015}.
	This method makes it possible to estimate the properties of many M dwarfs relatively easily with sufficient precision, and enables us to efficiently select M dwarfs that are suitable for large exoplanet surveys.
	We also provide the spectral types of our sample.
	
	The rest of the paper is organized as follows.
	In section \ref{sec:data}, we describe our sample and observations.
	Section \ref{sec:analysis} details our analysis.
	Our results are summarized in section \ref{sec:result}.
	We discuss the validity of our results, and compare them with the other studies in section \ref{sec:discussion}.
	We conclude the paper in section \ref{sec:conclusion} with a summary of our study.

\section{Data \label{sec:data}}
	
	\subsection{Target sample}
	Our target sample was drawn from the LSPM catalog \citep{Lepine2005}, which comprises stars with annual proper motions larger than $0\farcs15$.
	We attempted to exclude binaries and multiple-star systems from the sample for our spectroscopic characterizations.
	We then used the SIMBAD database to identify eclipsing binaries, double or multiple stars, and spectroscopic binaries.
	Based on the Wasignton Double Star catalogue (WDS; Mason et al. 2001, 2014), we also avoided the stars from which companion candidates are located at the projected separations smaller than 1000 AU\footnote{This work made use of the spectra which were obtained in the IRD target selection. They applied 1000 AU threshold for their sample selection. See IRD proposal (https://www.naoj.org/Science/SACM/Senryaku/IRD\_180520235849.pdf) for detail.}.
	In spite of these attempts, our sample still includes some known multiple-star systems, which are marked in table \ref{tbl_result} if they are found in the WDS catalogue. 
	
\begin{figure}
		\centering
		\includegraphics[width=\columnwidth]{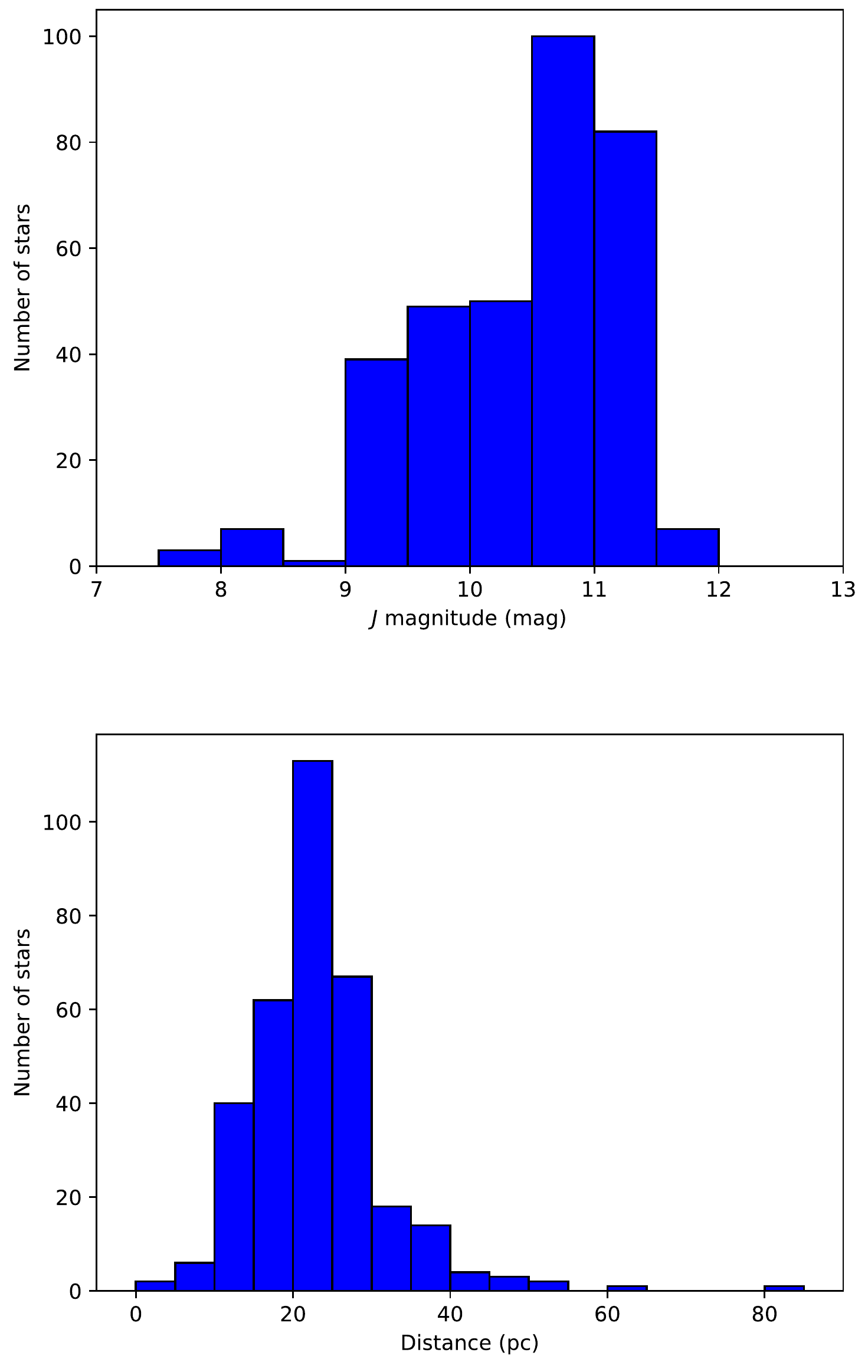}
		\caption{
			Histograms of our samples. $J$-band magnitudes (top), distances (bottom).
			\label{fig_object_hist}
		}
	\end{figure}

	Our sample consists of 338 M dwarfs.
	Their brightness and parallaxes (plx) lie in the ranges $7<J<11.5$ and $0\farcs02<\rm{plx}<0\farcs13$.
	The $J$ magnitudes were drawn from the Two Micron All Sky Survey (2MASS, \cite{Cutri2003}), and parallaxes were drawn from Gaia DR2 \citep{Gaia2018} or MEarth observations \citep{Dittmann2014}\footnote{We also partially used the Hipparcos \citep{vanLeeuwen2007} parallaxes as for the selection of targets in this study, but those parallaxes were not finally adopted in the analysis below.}.
	The fraction of stars with Gaia DR2 parallaxes is 92\%.
	Figure \ref{fig_object_hist} shows histograms of our sample.

	\subsection{Observations \label{sec:data_obs}}
		The majority ($N=249$) of the spectra of our sample were obtained using the Kyoto Okayama Optical Low-dispersion Spectrograph (KOOLS; \cite{Yoshida2005}) mounted on the 188 cm telescope at Okayama Astrophysical Observatory (OAO) and the Bench spectrograph \citep{Bershady2008,Knezek2010} mounted on the 3.5-m telescope of the WYIN observatory\footnote{The WIYN Observatory is a joint facility of the University of Wisconsin-Madison, Indiana University, the National Optical Astronomy Observatory and the University of Missouri.}. The KOOLS observations were conducted from August 2014 to April 2016, and the Bench Spectrograph observations were performed in November 2016 and May 2017. In addition, only 21 samples were observed with the Medium And Low-dispersion Long-slit Spectrograph (MALLS; \cite{Ozaki2005}) mounted on the 2-m telescope at Nishi-Harima Astronomical Observatory (NHAO) from December 2016 to July 2017 and the Calar Alto Faint Object Spectrograph (CAFOS; \cite{Meisenheimer1994}) mounted on the 2.2 m telescope at Calar Alto Observatory (CAO)\footnote{Based on observations collected at the Centro Astron\'{o}mico Hispano Alem\'{a}n (CAHA) at Calar Alto, operated jointly by the Max-Planck Institut f\"{u}r Astronomie and the Instituto de Astrof\'{i}sica de Andaluc\'{i}a (CSIC).} in October 2016. Finally, the Dual Imaging Spectrograph (DIS)\footnote{https://www.apo.nmsu.edu/arc35m/Instruments/DIS/} mounted on the ARC 3.5-m telescope at Apache Point Observatory (APO)\footnote{Based on observations obtained with the Apache Point Observatory 3.5-meter telescope, which is owned and operated by the Astrophysical Research Consortium.} from November 2016 to March 2018 were utilized to take the spectra of 70 stars.
		KOOLS was upgraded to KOOLS-IFU \citep{Matsubayashi2019} in February 2016, but the spectrograph itself did not change.
		Slit widths were set to 1$\farcs$0 for KOOLS, 1$\farcs$2 for MALLS, 1$\farcs$2 for CAFOS, and 1$\farcs$5 for DIS, respectively.
		The KOOLS-IFU have fiber arrays on their focal planes, whose widths were set to 1$\farcs$8 for KOOLS-IFU.
		We adopted $\nabla$Pak\footnote{Pronounced "GradPak."} \citep{Eigenbrot2018}, which have the rectangular arrays with the rows of fibers whose field-of-views (FoVs) varies from 1$\farcs$87 to 5$\farcs$62, to inject the lights from our targets into to the Bench spectrograph.
		We then used the smallest-FoV fiber in our observations.
		Table \ref{tbl_inst} shows the details of the instruments we used, and our observations are summarized in table \ref{tbl_observationlog}.
		Our spectra have the typical signal-to-noise (SN) ratios at 7400 $\mathrm{\AA}$ of $\sim80$ for KOOLS, $\sim36$ for KOOLS-IFU, and $\sim100$ for the other instruments.
		The noise values are just based on the photon noise of each obtained spectrum.
		We note that the CCD of KOOLS had bad pixels in the range 6675--6715 \AA, and therefore we did not use this area in our analysis.
		\begin{table*}
			\tbl{Instruments and telescopes used in this work.\footnotemark[$*$] }{%
			\begin{tabular}{lccccc}
				\hline\noalign{\vskip3pt} 
				Instrument and telescope & Wavelength (\AA) & Disperser & Resolution ($R \sim \lambda/\Delta\lambda$) & Number of stars & Number of nights \\  [1pt] 
				\hline\noalign{\vskip3pt} 
					KOOLS/OAO 188 cm & 6400--7600 & VPH683 & 2000 & 91 & 17 \\
					KOOLS-IFU/OAO 188 cm & 6400--7600 & VPH683 & 2000 & 33 & 6 \\
					MALLS/NHAO 2 m & 5200--8200 & 300 l/mm & 1200 & 18 & 8 \\
					Bench spectrograph/WIYN 3.5 m & 5000--7900 & 600@10.1 & 1700 & 125 & 9 \\
					CAFOS/CAO 2.2 m & 4200--8300 & G-100 & 1700 & 3 & 2 \\
					DIS/APO 3.5 m & 4160--5430, 6100--7300 & B1200, R1200 & 10000 & 68 & 7 \\  [2pt] 
				\hline\noalign{\vskip3pt} 
			\end{tabular}} \label{tbl_inst}
			\begin{tabnote}
				\hangindent6pt\noindent
				\hbox to6pt{\footnotemark[$*$]\hss}\unskip%
				KOOLS-IFU and Bench spectrograph transfer object light to their spectrographs through fibers of widths 1$\farcs$8 (KOOLS-IFU) and 1$\farcs$87 (Bench spectrograph).
			\end{tabnote}
		\end{table*}
		
\begin{figure}
			\centering
			\includegraphics[width=\columnwidth]{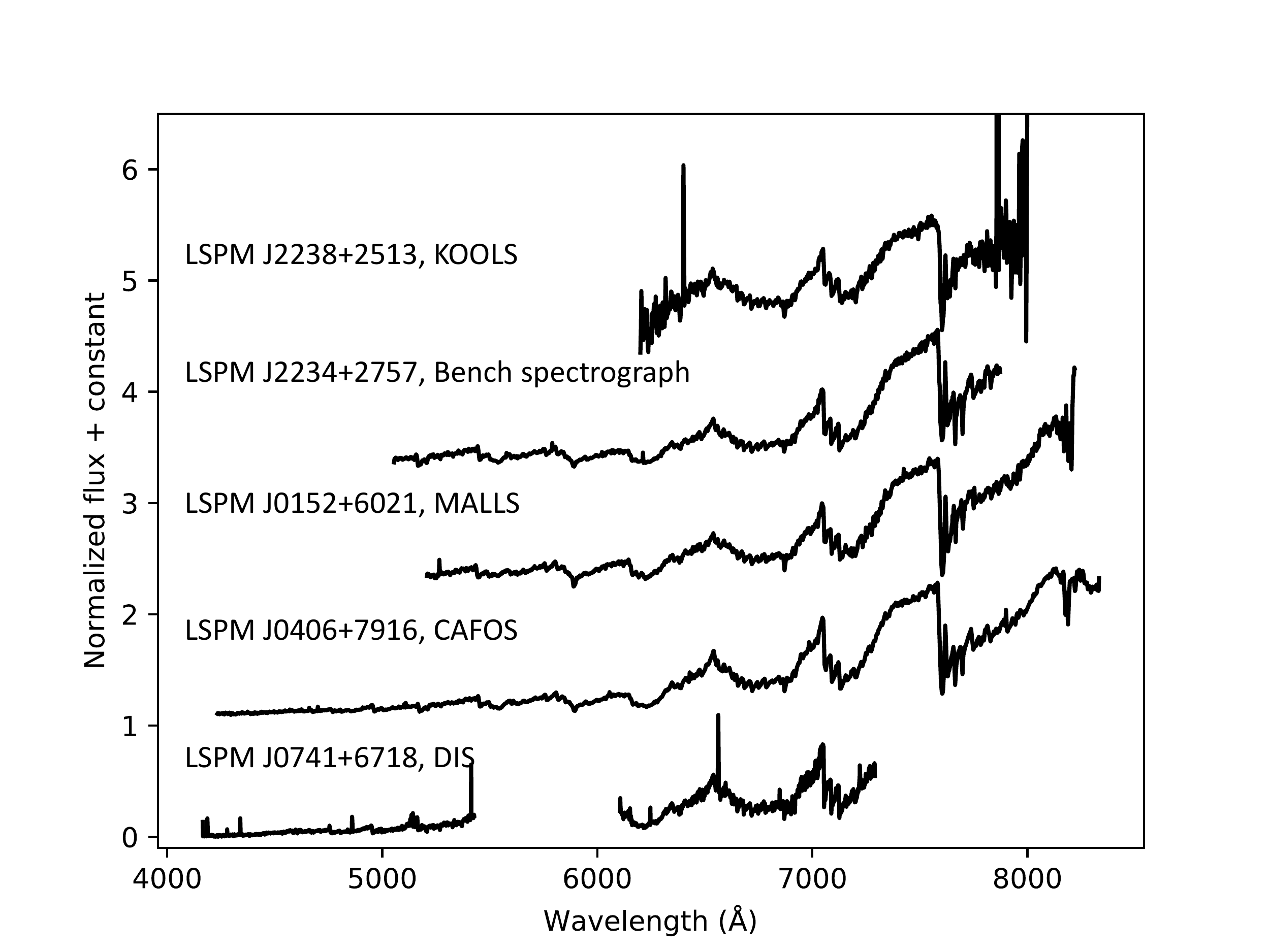}
			\caption{
				Examples of our samples.
				The top one is LSPM J2238+2513 obtained by KOOLS, the second from the top is LSPM J2234+2757 obtained by the Bench spectrograph, the third from the top is LSPM J0152+6021 obtained by MALLS, the fourth from the top is LSPM J0406+7916 obtained by CAFOS, and the bottom sample is LSPM J0741+6718 obtained by DIS.
				We note that the CCD of KOOLS had bad pixels in 6675--6715 \AA, which were excluded in this study.
				\label{fig_spectra}
			}
		\end{figure}

		We used IRAF\footnote{IRAF is distributed by the National Optical Astronomy Observatories, which is operated by the Association of Universities for Research in Astronomy, Inc. under cooperative agreement with the National Science Foundation, USA.} for the data reduction, which included bias subtraction, flat fielding, sky-background subtraction, aperture extraction, wavelength calibration, and instrumental response correction.
		We also used the data-reduction software\footnote{http://www.oao.nao.ac.jp/\~{}kools/man/man\_reduction.html} for KOOLS and the pipelines\footnote{http://www.kusastro.kyoto-u.ac.jp/\~{}kazuya/p-kools/reduction-201604/index.html} developed for KOOLS-IFU. 
		Our wavelength calibrations were based on a Fe-Ne-Ar lamp for KOOLS, KOOLS-IFU, and MALLS, a Cu-Ar lamp for Bench spectrograph, and a Hg-He-Rb lamp for CAFOS.
		On each night, we observed some spectrophotometric standards listed in \citet{Oke1990} to correct the instrumental responses of the telescopes and instruments.
		Examples of the reduced spectra are shown in Figure \ref{fig_spectra}.

	\subsection{Data from other literature}
		We also analyzed the data of \citet{Mann2015} to calibrate the stellar properties derived from our observed spectra.
		They obtained the 183 optical and near-infrared spectra of nearby K7--M7 single dwarfs obtained from the CONCH-SHELL catalog \citep{Gaidos2014} and \citet{Lepine2011}, using the SuperNova Integral Field Spectrograph (SNIFS, \cite{Aldering2002,Lantz2004}) mounted on the University of Hawaii 2.2 m telescope at Mauna Kea. 
		Their spectra covered the wavelength range 3200--9700 $\rm{\AA}$ with a wavelength resolution of $R\sim 1000$.
		The median distance of their sample was $\sim$10 pc, and their $[\rm{Fe/H}]$ ranges from $-0.60$ to $+0.53$.
		\citet{Mann2015} derived the $T_{\mathrm{eff}}$ of their sample by fitting the CFIST suite\footnote{https://phoenix.ens-lyon.fr/Grids/BT-Settl/} of the BT-Settl version of the PHOENIX model \citep{Allard2013} to the observed spectra, using the method detailed in \citet{Mann2013b}, and achieved a $T_{\mathrm{eff}}$ precision of $\sim$60 K.

\section{Analysis\label{sec:analysis}}
	In this section, we explain our method of determining stellar $T_{\mathrm{eff}}$, radii, and spectral types.
	The calculation of $T_{\mathrm{eff}}$ is based on synthetic spectra, and the radii are estimated from photometric magnitudes and the Gaia parallaxes, while we determined the spectral types from molecular absorption features.
	The $T_{\mathrm{eff}}$ are calibrated with the optical spectra of the M dwarfs of \citet{Mann2015}. 
	Our methods are based on those used by \citet{Mann2015} (and reference therein), who provide the well-calibrated $T_{\mathrm{eff}}$ estimates.
	We expand their approach to include different wavelength regimes, and note that our methodology is even applicable to datasets having relatively narrow wavelength coverage, such as KOOLS.

	\subsection{Temperature \label{sec:analysis_teff}}
		
		\subsubsection{Temperature from optical spectra \label{sec:analysis_teff_modelfit}}
			PHOENIX is a widely used software program to make synthetic model spectra for determining the $T_{\mathrm{eff}}$ of M dwarfs (e.g., Mann et al. 2013b; 2015).
			In our study, stellar $T_{\mathrm{eff}}$ were calculated by comparing our observed spectra with the latest PHOENIX-ACES model grid.
			The PHOENIX-ACES model grid\footnote{http://phoenix.astro.physik.uni-goettingen.de} that we used is detailed in \citet{Husser2013}.
			We found the model spectrum that minimized the following $\chi^2$ value:
			\begin{equation}
				\chi^2 = \frac{1}{N-1} \sum_i \Bigl\{ \frac{F_{\mathrm{obs}.i} - n\,F_{\mathrm{model},i}}{\sigma_i} \Bigr\}^2, \label{eq_chisq_spec}
			\end{equation}
			where $F_{\mathrm{obs}}$ is an observed spectrum, $F_{\mathrm{model}}$ is a PHOENIX model spectrum, $\sigma$ is the flux error and $n$ is the normalization coefficient.
			For the normalization factor $n$ in the above fit, we attempted a simple constant normalization, and a linear, quadratic, and cubic polynomial.
			We adopted constant normalization for $n$ in Equation (\ref{eq_chisq_spec}), due to the reason described in Appendix \ref{appedix_linearcorrection}.
			Briefly, constant normalization can best reproduce the calibrations of \cite{Mann2015} (see also Section \ref{sec:analysis_teff_linearcorrection}), so it can produce the most accurate $T_{\mathrm{eff}}$ estimations among our adopted normalizations.
			
			$T_{\mathrm{eff}}$, $[\rm{Fe/H}]$, and $\log{g}$ of the synthetic spectra range from 2300 to 4500 K in increments of 100 K, from $-$1.0 to 1.0 in increments of 0.1, and 4.5 or 5.0, respectively.
			We note that the original grid interval of the model for $\log{g}$ and metallicity were 0.5, which we interpolated linearly to obtain an interval of 0.1.
			A synthetic model spectrum was convolved with a Gaussian whose width corresponds to the wavelength resolution of each instrument.
			We discarded the wavelength ranges affected by telluric absorptions and the possible $\rm{H_\alpha}$ emissions.
			Additionally, we omitted the region of 6400--6600 $\mathrm{\AA}$, in which there is a TiO absorption that remains poorly modeled \citep{Reyle2011} and was omitted in the analysis of \citet{Mann2013b,Lepine2013}.
			One should note there is still the inaccuracy of TiO line lists, as seen in the comparison of high-resolution spectroscopy of M dwarfs to the modeled TiO spectra \citep{Hoeijmakers2015,McKemmish2019}.
			First, we calculated $\chi^2$ values for spaced model grid to get a rough global minimum.
			After determining the best-fit model from all models that are sampled with the $T_{\mathrm{eff}}$ interval of 100 K, we further explore a better fit around the initial best-fit model as below.  
			We selected two additional models: one has a higher $T_{\mathrm{eff}}$ than the best-fit model by 100 K, and the other has a lower $T_{\mathrm{eff}}$ by 100 K.
			For the stars observed with KOOLS or KOOLS-IFU, we also select two models, one of which has a higher $T_{\mathrm{eff}}$ than the best-fit model by 200 K, and the other has a lower $T_{\mathrm{eff}}$ by 200 K.
			Then, these models including the initial best-fit model (i.e, $T_{\mathrm{eff}}=$ best $-$ 100 K, best, best $+$ 100 K) are linearly interpolated, from which we further seek the final best-fit model spectrum.
			The metallicity and $log{g}$ were also optimized again, with the same interpolations as described above. 
			Next, we generated 1000 mock data sets by adding artificial noise to the best-fit spectrum, and repeated our model fit procedure.
			The metallicity and $\log{g}$ are fixed at the best-fitted values.
			In this step, the artificial noise was extracted from a Gaussian distribution whose standard deviation is equal to the value large enough to scale the reduced $\chi^2$ of the best fit to unity.
			Thus, we repeated the optimizations for the 1000 mock data sets, and we calculate the average and standard deviation of 1000 $T_{\mathrm{eff}}$ optimizations, which are accepted as the final estimation of a stellar $T_{\mathrm{eff}}$ and its error value.

		\subsubsection{Calibration using the literature data \label{sec:analysis_teff_linearcorrection}}
\begin{figure}
				\centering
				\includegraphics[width=\columnwidth]{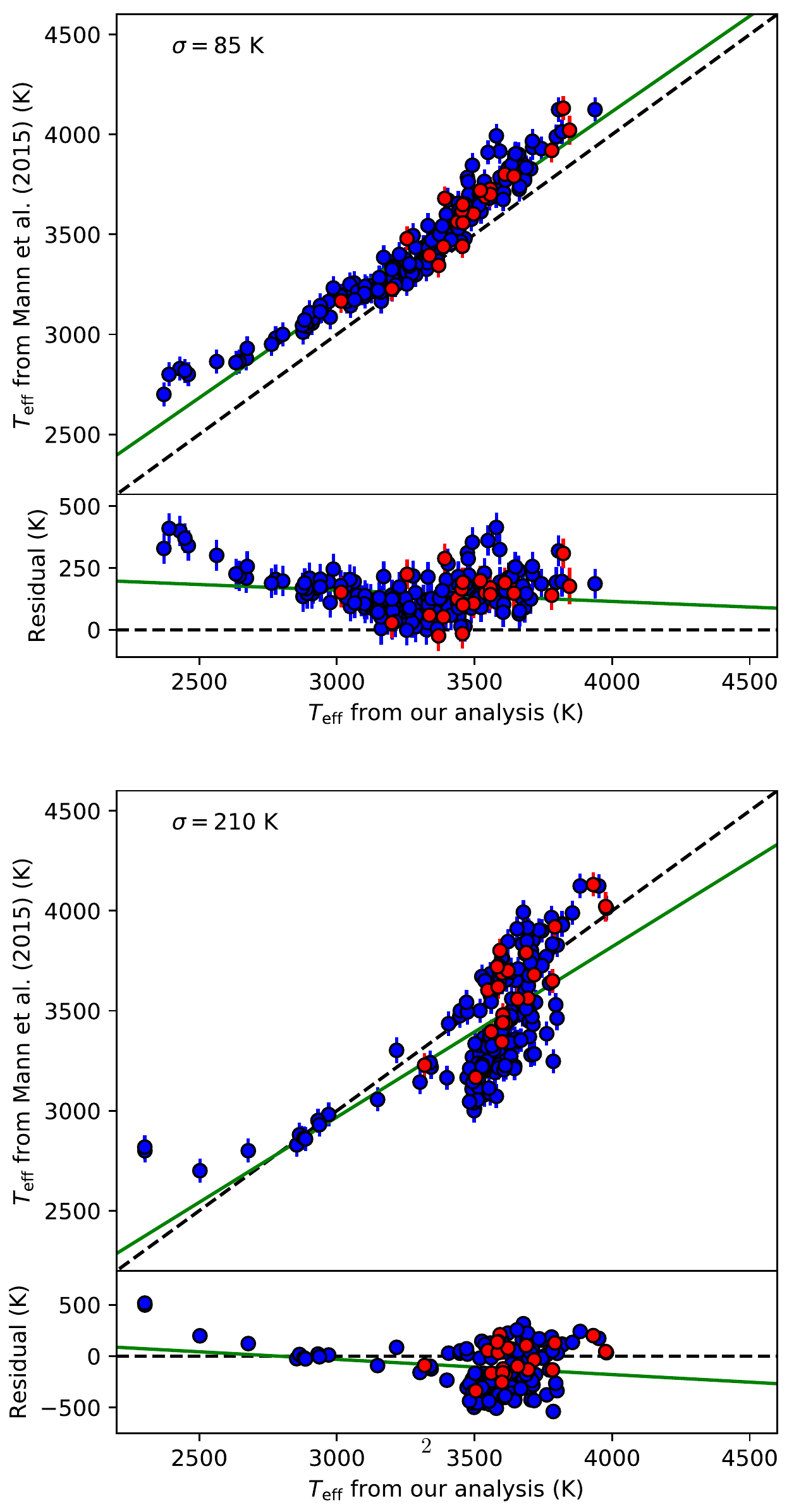}
				\caption{
					Comparison of our method with \citet{Mann2015}.
					Top panel is the result of 5000 -- 8000 \AA, and bottom is that of 6600 -- 7500 \AA.
					Green lines in the figure represent linear functions fitted to the data.
					Red points represent stars which have interferometric observations \citep{Lane2001, Segransan2003, Berger2006, Kervella2008, Demory2009, vonBraun2011, vonBraun2012, vonBraun2014, Boyajian2012}.
					\label{fig_teff_comparison_mann}
				}
			\end{figure}

			We tested whether the wavelength coverage of each instrument is adequate to determine stellar $T_{\mathrm{eff}}$ (see table \ref{tbl_inst}).
			For this test, we fitted the PHOENIX model spectra to the data of \citet{Mann2015} in the same way as described in the previous section.
			This analysis was applied to the data of \citet{Mann2015} under two different conditions: we first analyzed the wavelength range of 6600--7500 \AA\ that corresponds to KOOLS, and we next analyzed the wavelength range 5000--8000 \AA\ corresponding to the other instruments.
			We compared the $T_{\mathrm{eff}}$ thus obtained from the optical wavelength ranges, $T_{\rm{optical}}$, with those obtained from the optical and near-infrared ranges by \citet{Mann2015}, $T_{\rm{Mann}}$, and expressed $T_{\rm{Mann}}$ as a linear function of $T_{\rm{optical}}$:
			\begin{equation}
				T_{\rm{Mann}} = a\,T_{\rm{optical}} + b. \label{eq_linearcorrection}
			\end{equation}
			Figure \ref{fig_teff_comparison_mann} shows a comparison between our derivations and the results of \citet{Mann2015}.
			Our results do not follow one-to-one relation, which is probably due to differences in PHOENIX versions and differences in the wavelength range.
			As can be seen in Figure \ref{fig_teff_comparison_mann}, the estimation of $T_{\mathrm{eff}}$ becomes more accurate if we use a wavelength range of 5000--8000 \AA.
			The residuals of the best-fit function have the standard deviations corresponding to 210 K for 6600--7500 $\mathrm{\AA}$ and 85 K for 5000--8000 \AA.
			We found that our linear correction for the fit of 5000--8000 \AA\ has relatively large deviations ($>2\sigma$) in the low temperature region with $T_{\mathrm{eff}}<2600$ K.
			For the fit of 5000--8000 \AA\ the removal of stars with $T_{\mathrm{eff}}<2600$ K decreases the standard deviation of the residuals to be 82 K, however which is comparable to the case of no removal.
			
\begin{figure}
				\centering
				\includegraphics[width=\columnwidth]{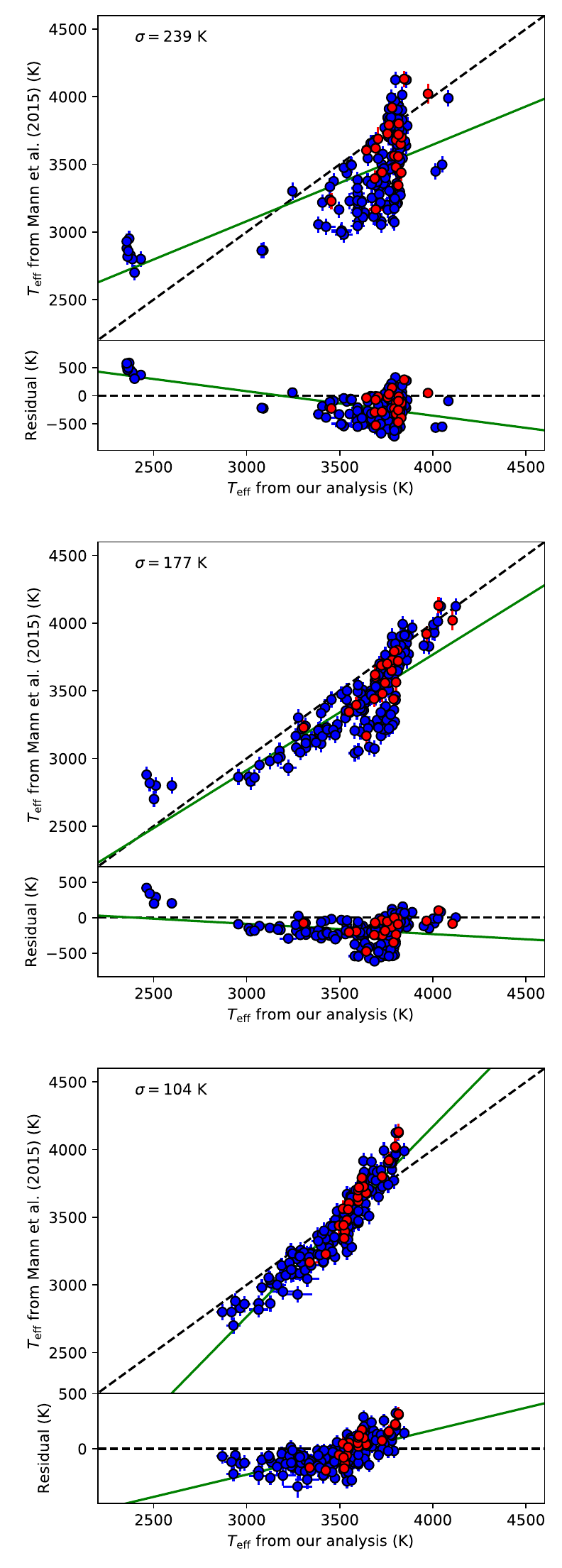}
				\caption{
					Same as Figure \ref{fig_teff_comparison_mann}, but temperatures from synthetic spectra using molecular absorption bands listed in table \ref{tbl_specband}.
					Top panel is TiO band1, middle is TiO band2, and bottom is VO band.
					\label{fig_teff_comparison_narrowband}
				}
			\end{figure}

\begin{figure}
				\centering
				\includegraphics[width=\columnwidth]{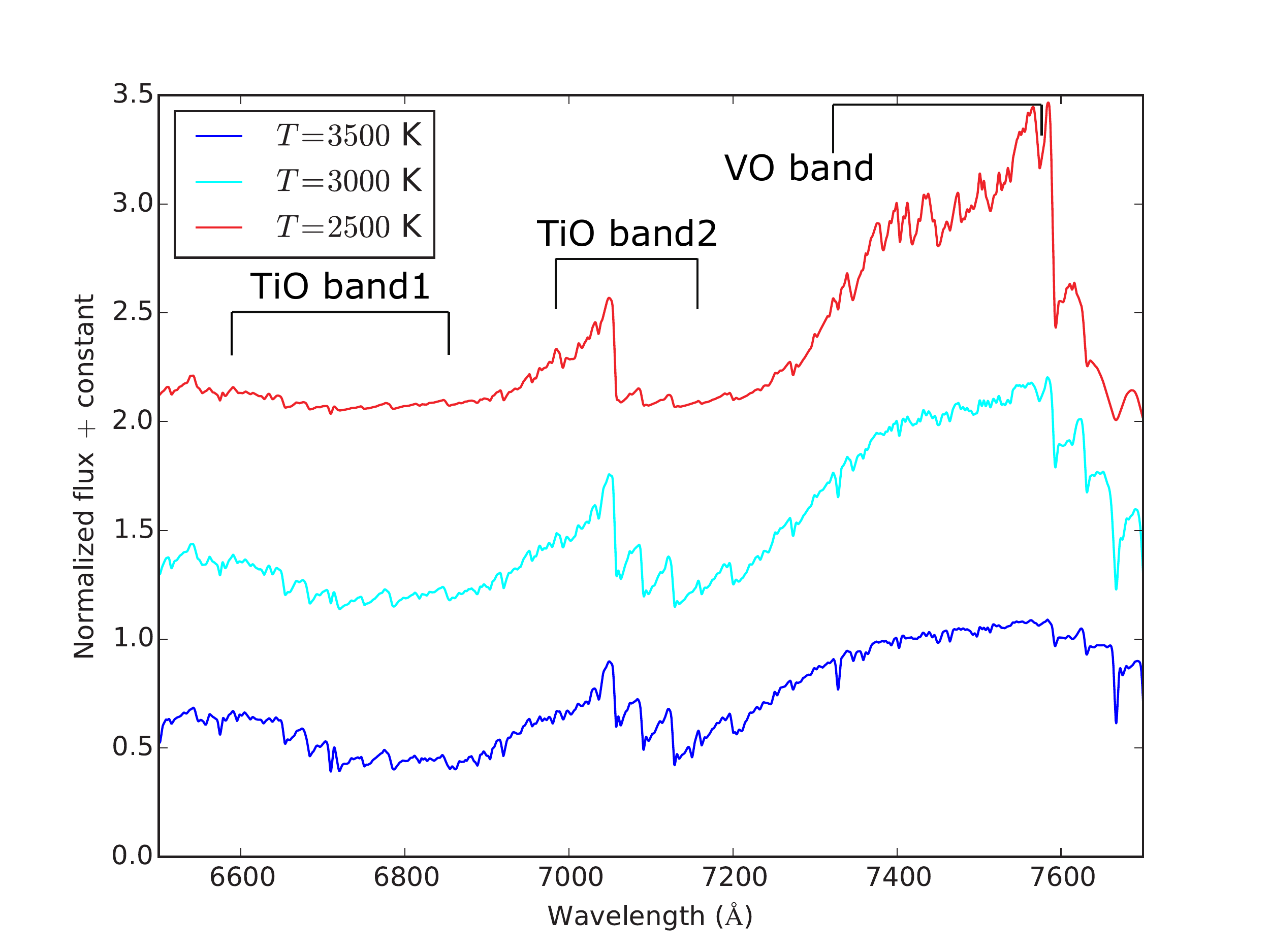}
				\caption{
					Synthetic spectra of \citet{Husser2013}, which we used for fits to estimate stellar $T_{\mathrm{eff}}$.
					The metallicity was fixed as $\rm{[Fe/H]}=0.0$ in this plot.
					The VO band is a region in 7320--7570 \AA.
					\label{fig_model_teffdependence}
				}
			\end{figure}

			The narrow wavelength coverage of KOOLS and KOOLS-IFU prevents us from accurately determining stellar $T_{\mathrm{eff}}$ (see top panel of Figure \ref{fig_teff_comparison_mann}).
			To improve the precision of our $T_{\mathrm{eff}}$ estimation with the wavelength range of 6600--7500 \AA, we focus on the wavelength ranges of the three molecular absorption bands listed in table \ref{tbl_specband} rather than using a whole wavelength range.
			These three regions are less affected by telluric absorptions in the spectra obtained by KOOLS or KOOLS-IFU (see Figure \ref{fig_model_teffdependence}).
			We reanalyzed the data of \citet{Mann2015} using these three bands, and the results for each band are shown in Figure \ref{fig_teff_comparison_narrowband}.
			The green lines in Figure \ref{fig_teff_comparison_narrowband} represent the linear functions that were fit to the relations between our re-analysis of the data in \citet{Mann2015} and the original analysis of \citet{Mann2015}.
			The VO band was the most sensitive to $T_{\mathrm{eff}}$ (see Figure \ref{fig_model_teffdependence}).
			The best-fit linear functions have the residual scatters of 239 K for the fit using TiO band1, 177 K for the fit using TiO band2, and 103 K for the fit using the VO band.
			Therefore, we used the VO band for the data obtained by KOOLS in calculating the $T_{\mathrm{eff}}$.
			\begin{table}
				\tbl{Molecular absorption bands used in synthetic spectra fitting for KOOLS spectra.}{
				\begin{tabular}{cc}
					\hline\noalign{\vskip3pt} 
					Band & Wavelength range (\AA)  \\  [2pt] 
					\hline\noalign{\vskip3pt} 
						TiO band1 & 6600--6850 \\
						TiO band2 & 6960--7160 \\
						VO band & 7320--7570 \\  [2pt]
					\hline\noalign{\vskip3pt} 
				\end{tabular}} \label{tbl_specband}
			\end{table}

		\subsubsection{Estimation of temperature for our data}
\begin{figure}
				\centering
				\includegraphics[width=\columnwidth]{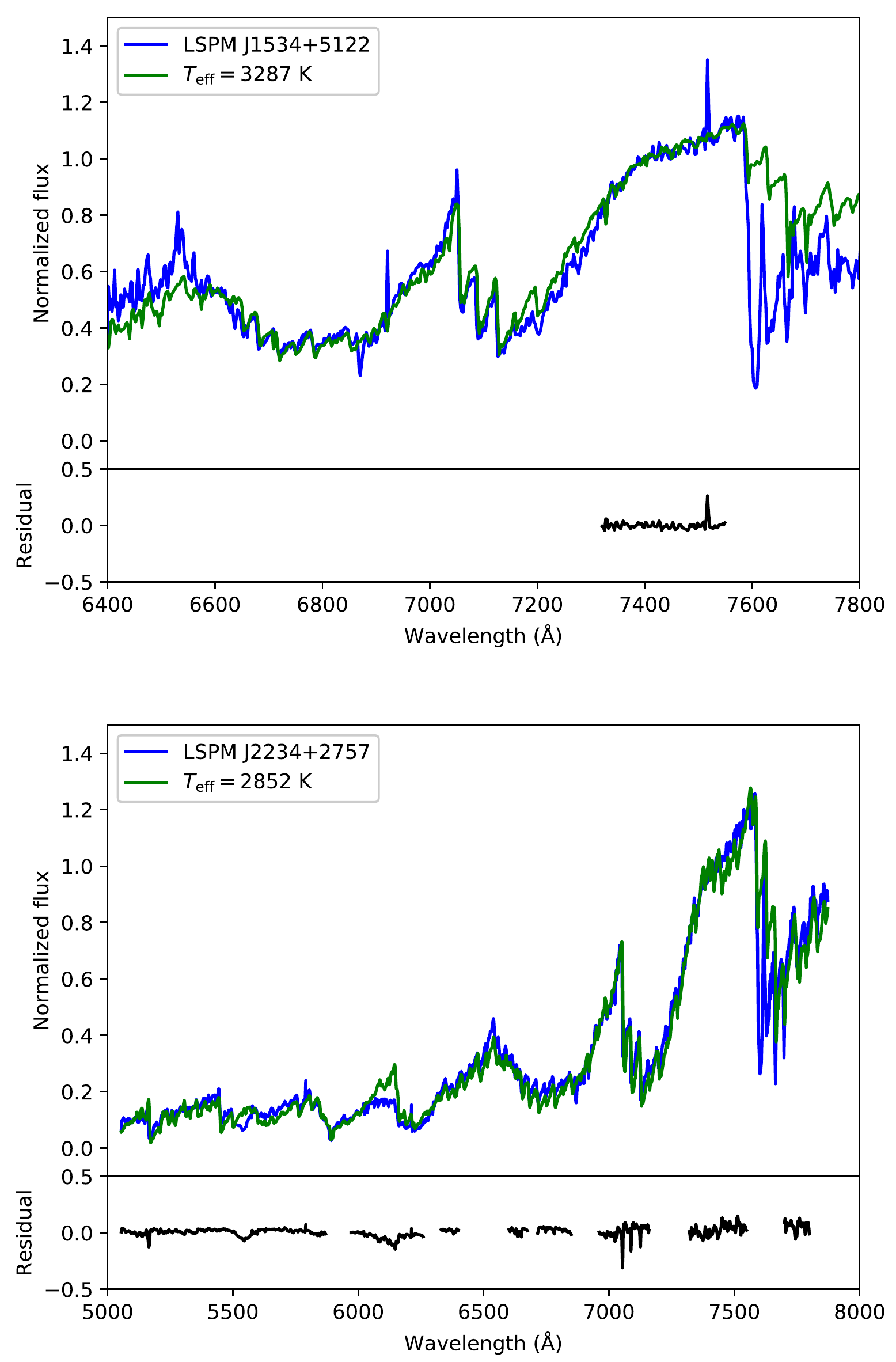}
				\caption{
					Examples of optical spectra (blue) and best-fit synthetic spectra (green).
					Top panel is LSPM J1534+5122 obtained by KOOLS, and bottom panel is LSPM J2234+2757 obtained by Bench spectrograph.
					The wavelength ranges of residuals in both panels correspond to the wavelengths that were utilized in our fitting analyses.
					\label{fig_fitexamples}
				}
			\end{figure}

			We fitted PHOENIX model spectra to our data, and estimated their $T_{\mathrm{eff}}$.
			As mentioned above, the fit used the VO band for the data obtained by KOOLS and full spectral ranges for the data obtained by the other instruments.
			Figure \ref{fig_fitexamples} shows examples of optical spectra and best-fit synthetic spectra.
			
			We then applied linear corrections derived by comparison with \citet{Mann2015} (Equation \ref{eq_linearcorrection}).
			The coefficients of linear corrections are listed in table \ref{tbl_linearcorrection}.
			We implement the error of linear correction into a total $T_{\mathrm{eff}}$ uncertainty by calculating $\sigma_{\rm{Total}} = \sqrt{\sigma^2_{\rm{fit}} + \sigma^2_{\rm{corr}}}$, where $\sigma_{\rm{fit}}$ is the error of fit and $\sigma_{\rm{corr}}$ is the error of linear correction (see Table \ref{tbl_linearcorrection}).
			The standard deviations of residuals in the linear-function fitting (see section \ref{sec:analysis_teff_linearcorrection}) were taken to be the errors of linear corrections.
			We did not calculate the $T_{\mathrm{eff}}$ of the stars observed by DIS because their wavelength range was not suitable for our analysis.
			\begin{table}
				\tbl{Coefficients of linear corrections (Equation \ref{eq_linearcorrection}) for each band.}{%
				\begin{tabular}{cccc}
					\hline\noalign{\vskip3pt} 
					Band & $a$ & $b$ & $\sigma_{\mathrm{corr}}$ (K)  \\  [2pt] 
					\hline\noalign{\vskip3pt} 
						VO band & $1.405$ & $-1449.9$ & 103 \\
						5000--8000 \AA & $0.954$ & $296.6$ &  85 \\ [2pt] 
					\hline\noalign{\vskip3pt} 
				\end{tabular}} \label{tbl_linearcorrection}
			\end{table}

	\subsection{Radii from Spectral Energy Distribution (SED) \label{sec:analysis_sed}}
		We determined the stellar radii and $T_{\mathrm{eff}}$ from the SED.
		The SED is a distribution of the flux density in wavelength or frequency units, and is used to characterize stellar parameters (e.g., $T_{\mathrm{eff}}$, radius, and interstellar extinction).
		Such stellar properties can be calculated by comparing photometric magnitudes with synthetic stellar spectra.
		
		For each star, we retrieved their $V$ magnitudes from the fourth U.S. Naval Observatory CCD Astrograph Catalog (UCAC4, \cite{Zacharias2012}) and \citet{Lepine2005}; Gaia $G$, $G_{\mathrm{BP}}$, and $G_{\mathrm{RP}}$ magnitudes from Gaia DR2 \citep{Gaia2018}; $J$, $H$, and $K$ magnitudes from 2MASS \citep{Cutri2003,Skrutskie2006}; and $W1$, $W2$, $W3$, and $W4$ magnitudes from the Wide-field Infrared Survey Explorer (WISE, \cite{Wright2010}).
		The photometric measurements used for our SED analysis are provided in table \ref{tbl_stellarparameters1} and \ref{tbl_stellarparameters2}. 
		All photometric systems and their effective wavelengths that we used are shown in table \ref{tbl_photomag}.
		We compiled the flux zero points for $V$ band from \citet{MannBraun2015}, Gaia bands from \citet{Evans2018}, $JHK$ bands from \citet{Cohen2003}, and Wise bands from \citet{Wright2010}.
		The instrument response curves were drawn from \citet{MannBraun2015} for $V$ band , \citet{Evans2018} for Gaia bands, \citet{Cohen2003} for $JHK$ bands, and \citet{Jarrett2011} for Wise bands.
		
		We used the PHOENIX BT-SETTL version \citep{Allard2011} in SED analysis. 
		We note that the model grid is different from that in the model fitting for optical spectra.
		The synthetic spectra of \citet{Husser2013} cover the wavelength range from 0.5 $\rm{\mu m}$ to 5.5 $\rm{\mu m}$.
		However, the effective wavelengths of the $W3$ and $W4$ are longer than 5.5 $\rm{\mu m}$.
		Therefore, we could not use the synthetic spectra of \citet{Husser2013} in the SED analysis.
		
\begin{figure}
			\centering
			\includegraphics[width=\columnwidth]{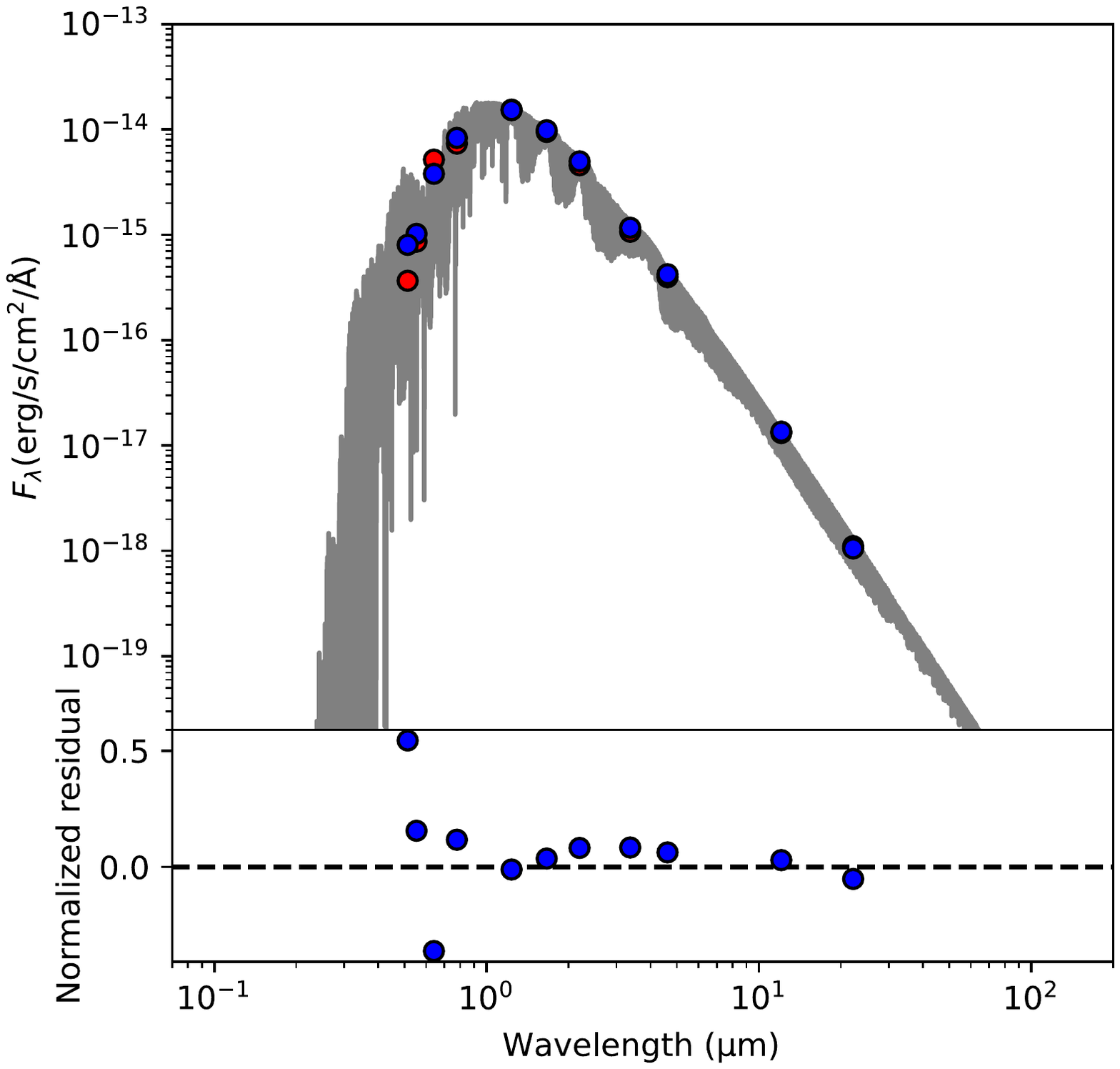}
			\caption{
				Example of spectral energy distribution (SED). This is a plot of LSPM J1502+7527.
				Blue points represent photometry, red points represent photometry calculated from synthetic spectra and the gray line is the best-fitted synthetic spectrum in the SED calculation.
				Each axis uses the logarithmic scale. The $T_{\mathrm{eff}}$ of the synthetic spectrum is 2837 K, and that from optical spectrum is 2802 K.
				\label{fig_sedfit}
			}
		\end{figure}

		The synthetic flux of model spectrum, $f_{\rm{model}}$, was calculated using the following formula:
		\begin{equation}
			f_{\rm{model}} = \frac{\int f(\lambda)S(\lambda)d\lambda}{\int S(\lambda)d\lambda},
		\end{equation}
		where $f(\lambda)$ is the PHOENIX spectrum as a function of the wavelength, and $S(\lambda)$ is a response curve of each band.
		The flux density of the PHOENIX model spectrum is the energy emitted per second, per wavelength, and per unit area on the stellar surface.
		Therefore, we multiplied it by $(R_*/d)^2$ to obtain the energy received per unit area on Earth.
		$R_*$ was obtained by minimizing $\chi^2$ as
		\begin{eqnarray}
			&\chi^2 = \displaystyle\frac{1}{N-1}\sum_i{\left\{\frac{f_{\mathrm{obs},i}-(R_*/d)^2f_{\mathrm{model},i}}{\sigma_i}\right\}}^2, \nonumber \\
			&R_* = d \sqrt{\displaystyle\frac{\sum_i\displaystyle\frac{f_{\rm{obs},i}f_{\rm{model},i}}{\sigma_i^2}}{\sum_i\displaystyle\frac{f_{\rm{model},i}^2}{\sigma_i^2}}}, \nonumber
		\end{eqnarray}
		where $d$ is the distance to the star, $f_{\rm{obs}}$ is the observed photometric flux, and $\sigma$ is the error in $f_{\rm{obs}}$.
		Stellar distances were taken from Gaia DR2 \citep{Gaia2018} and \citet{Dittmann2014}.
		The $T_{\mathrm{eff}}$ of the PHOENIX model grids range from 2300 to 4500 K. The surface gravity was set as 4.5 or 5.0, and the metallicity, $[\rm{M/H}]$, was fixed as 0.0.
		At first, we computed $\chi^2$ to get a rough minimum and then we interpolated $f_{\rm{model}}$ with $T_{\mathrm{eff}}$ to explore the global minimum around the initial rough minimum $\chi^2$.
		In this analysis, we finalize the optimizations based on the same manner as used in the analysis of our optical spectra (see section \ref{sec:analysis_teff}).
		Namely, we applied the Monte Carlo simulations to the photometic magnitudes and their scaled errors, and the best-estimate of $T_{\mathrm{eff}}$ for each target was represented by the median value of a simulated distribution, whose standard deviation was adopted as the error.
		Figure \ref{fig_sedfit} shows an example of SED fitting.
		\begin{table}
			\tbl{Photometric systems, effective wavelengths, and reference.}{%
			\begin{tabular}{lcc}
				\hline\noalign{\vskip3pt} 
				Band & Effective wavelength $(\mathrm{\mu m})$ & Reference \\  [2pt] 
				\hline\noalign{\vskip3pt} 
					$V$ & 0.5477 & UCAC4 and LSPM \\
					$G$ & 0.6405 & Gaia DR2 \\
					$G_{\mathrm{BP}}$ & 0.5131 & Gaia DR2 \\
					$G_{\mathrm{RP}}$ & 0.7778 & Gaia DR2 \\
					$J$ & 1.235 & 2MASS \\
					$H$ & 1.662 & 2MASS \\
					$K$ &  2.195 & 2MASS \\
					$W1$ & 3.368 & WISE \\
					$W2$ & 4.618 & WISE \\
					$W3$ & 12.082 & WISE \\
					$W4$ & 22.194 & WISE \\  [2pt]
				\hline\noalign{\vskip3pt} 
			\end{tabular}} \label{tbl_photomag}
		\end{table}

	\subsection{Spectral type}
		In this section, we describe our method of determining spectral types.
		We use spectral indices to derive spectral types.
		Spectral indices are widely used for spectral typing in low-resolution spectra.
		The indices, $i$, are based on molecular absorptions (e.g., \cite{Kirkpatrick1991,Reid1995}), and calculated as the ratio of flux in the following equation:
		\begin{equation}
			i = \frac{F(\Delta\lambda_{\rm{num}})}{F(\Delta\lambda_{\rm{den}})},
		\end{equation}
		where $\Delta\lambda_{\rm{num}}$ and $\Delta\lambda_{\rm{den}}$ are wavelength regions in numerator and denominator and $F(\Delta\lambda_{\rm{num}})$ and $F(\Delta\lambda_{\rm{den}})$ are the mean fluxes in the given wavelength ranges, $\Delta\lambda_{\rm{num}}$ and $\Delta\lambda_{\rm{den}}$, respectively.
		In general, either $\Delta\lambda_{\rm{num}}$ or $\Delta\lambda_{\rm{den}}$ is set to a continuum region.
		Our adopted indices and their wavelength ranges ($\Delta\lambda_{\rm{num}}$, $\Delta\lambda_{\rm{den}}$) are shown in table \ref{tbl_index}.
		\begin{table*}
			\tbl{Spectral-type indices and their wavelength that are used in this study. Indices were calculated as $i = F_{\rm{num}}/F_{\rm{den}}$. The regions near 7040, 7400, and 7510 \AA\ are continuum regions.}{
			\begin{tabular}{lccl}
				\hline\noalign{\vskip3pt} 
				Index & $\Delta\lambda_{\rm{num}} $ (\AA) & $\Delta\lambda_{\rm{den}} $ (\AA) & Reference  \\  [2pt] 
				\hline\noalign{\vskip3pt} 
					CaH2 & 6814--6846 & 7042--7046 & \citet{Reid1995} \\
					Ratio A (CaH) & 7020--7050 & 6960--6990 & \citet{Kirkpatrick1991} \\
					CaH Narr & 7044--7049 & 6972.5--6977.5 & \citet{Shkolnik2009} \\
					TiO3 & 7092--7097 & 7079--7084 & \citet{Reid1995} \\
					TiO5 & 7126--7135 & 7042--7046 & \citet{Reid1995} \\
					VO & $\sum$ 7385--7395, 7505--7515 & 7445--7460 & \citet{Kirkpatrick1995}  \\  [2pt] 
				\hline\noalign{\vskip3pt}
			\end{tabular}} \label{tbl_index}
		\end{table*}
		
\begin{figure*}
			\centering
			\includegraphics[width=16cm]{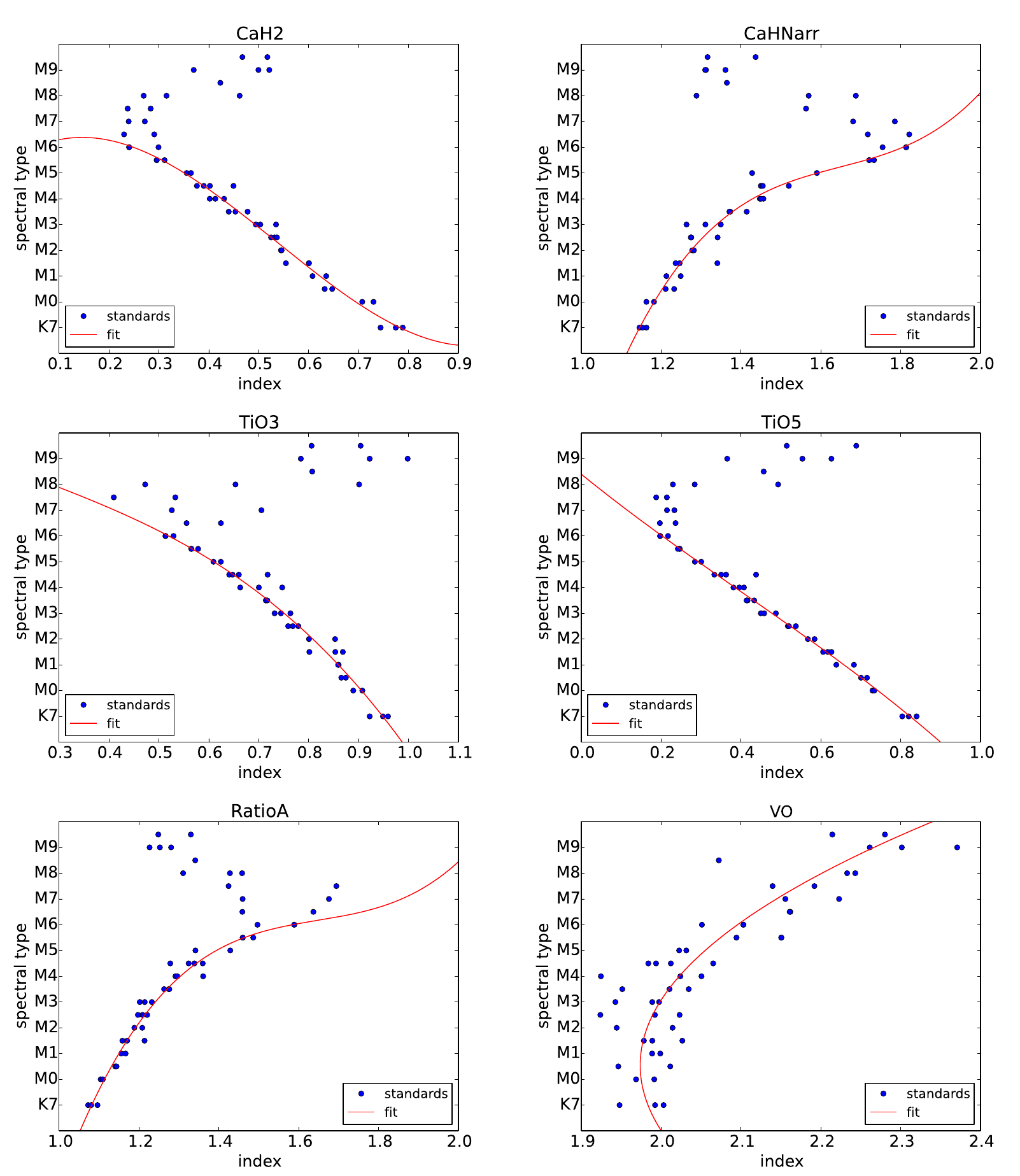}
			\caption{
				Relations between index values and spectral types (SpTs).
				Blue points represent the spectral standard stars listed in table \ref{tbl_sptstandards}, and red lines are the polynominal functions fitted to the index-SpT relations.
				\label{fig_sptindex}
			}
		\end{figure*}

		First, we derived relations between indices and spectral types, based on the spectral standard stars that were taken from the Palomar/Michigan State University survey\footnote{https://www.stsci.edu/\~{}inr/pmsu.html} (PMSU, \cite{Reid1995,Hawley1996}), \citet{Henry1990}, Kirkpatrick et al. (1991; 1999), \citet{Reid1999}, and Gizis et al. (2000a; 2000b).
		The spectral standard stars are summarized in table \ref{tbl_sptstandards}.
		We calculated each index value for the spectral standard stars, and fitted cubic polynomials to the relations between the spectral types and the index values. 
		Then, the spectral types, Spt, were expressed as cubic polynomials of each index as shown in the following equation:
		\begin{equation}
			{\rm Spt} = a\,i^3 + b\,i^2 + ci + d. \label{eq_spt}
		\end{equation}
		The coefficients of Equation (\ref{eq_spt}) are summarized in table \ref{tbl_sptcoef}.
		As can be seen in Figure \ref{fig_sptindex}, the spectral indices of CaH2, Ratio A, CaH Narr, TiO3, and TiO5 have similar values in early-M and late-M types; these indices are valid for dwarfs with spectral types from K7 V to M6 V.
		
		We considered the VO index to judge whether our samples were early-type stars or late-type stars. 
		VO is not sensitive to early-type stars, but it is sensitive to late-type stars (see Figure \ref{fig_sptindex}).
		We note that the VO index is slightly modified according to the spectrum of PMSU.
		The original definition of the VO index \citep{Kirkpatrick1995} is
		\begin{eqnarray}
			& {\rm VO} = \displaystyle\frac{a\,F_{7350\mathchar`-7400}+b\,F_{7510\mathchar`-7560}}{F_{7420\mathchar`-7470}}, \nonumber \\
			& a=0.5625,\ b=0.4375. \nonumber
		\end{eqnarray}
		However, there were some data in PMSU that did not have spectra beyond 7550 \AA.
		Therefore, we modified the definition of VO as follows:
		\begin{equation}
			{\rm VO} = \frac{F_{7385\mathchar`-7395}+F_{7505\mathchar`-7515}}{F_{7445\mathchar`-7460}}.
		\end{equation}
		The wavelength range of DIS is 6100--7300 \AA, and we did not consider the VO index for the spectra obtained by DIS.
		As a result, the spectral types of DIS data may be earlier than they actually are.
		We also define that the VO index is represented by a quadratic polynomial as follows in deriving the relation between the spectral type and the index:
		\begin{eqnarray}
			& {\rm VO} = a\,{\rm Spt}^2 + b\,{\rm Spt} + c \label{eq_vospt} \\
			& a = 4.093\times10^{-3}, \nonumber \\
			& b = -4.456\times10^{-3}, \nonumber \\
			& c = 1.975. \nonumber
		\end{eqnarray}
		\begin{table}
			\tbl{Coefficients of Equation (\ref{eq_spt}).}{
			\begin{tabular}{lcccc}
				\hline\noalign{\vskip3pt} 
				Index & $a$ & $b$ & $c$ & $d$  \\  [2pt]
				\hline\noalign{\vskip3pt} 
					CaH2 &$34.39$ &$-55.22$ &$13.96$ &$5.415$ \\
					Ratio A & $31.41$ & $-155.3$ & $258.6$ & $-138.8$ \\
					CaH Narr & $33.50$ & $-165.5$ & $276.7$ & $-151.2$ \\
					TiO3 &$-12.12$ &$7.889$ &$-9.272$ &$10.26$ \\
					TiO5 &$-3.700$ &$4.478$ &$-12.59$ &$8.399$ \\ [2pt]
				\hline\noalign{\vskip3pt} 
			\end{tabular}} \label{tbl_sptcoef}
		\end{table}
		
		Second, we computed each index value for our samples, and applied Equation (\ref{eq_spt}) and (\ref{eq_vospt}) to them.
		If stars were judged to be early or middle types (K7--M6) from the VO index, the median of the spectral types derived from the five indices (CaH2, Ratio A, CaH Narr, TiO3, and TiO5) was adopted as their best spectral types, and the error was identical as the standard deviation.
		If stars turned out to be later types (M6--9), we adopted the spectral types obtained from the VO index.

\section{Result \label{sec:result}}
	\begin{figure*}
		\centering
		\includegraphics[width=16cm]{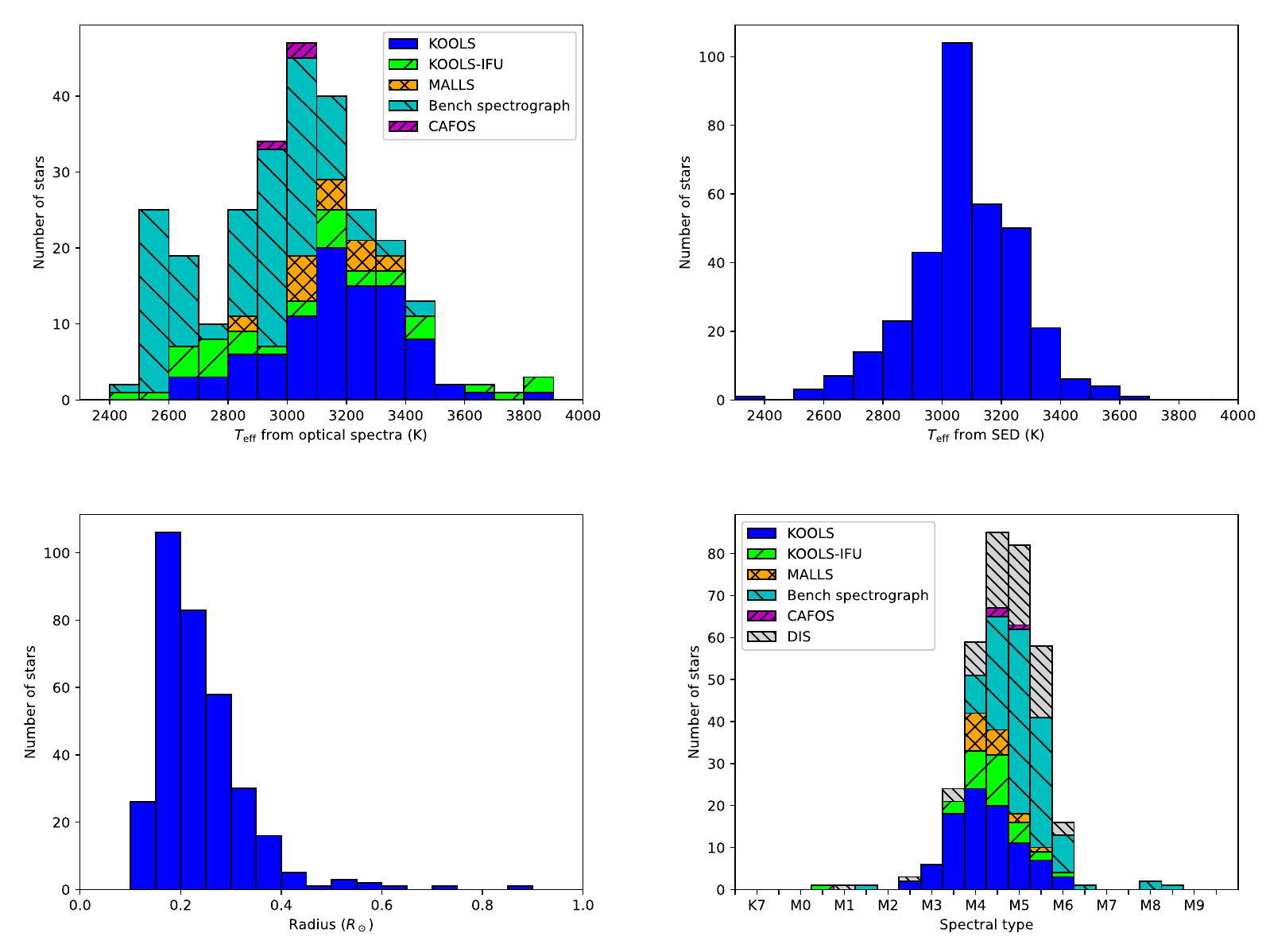}
		\caption{
			Histograms of our stellar-parameter derivations.
			Top left panel is a histogram of $T_{\mathrm{eff}}$ derived from optical spectra, the top right is that of $T_{\mathrm{eff}}$ derived from SED, the bottom left panel from the top is that of radii, and the bottom right panel is that of spectral types.
			\label{fig_result_histogram}
		}
	\end{figure*}

	\subsection{Temperature}
		We show the $T_{\mathrm{eff}}$ of our samples determined by using optical spectra in table \ref{tbl_result}, and a histogram of the $T_{\mathrm{eff}}$ at the top panel in Figure \ref{fig_result_histogram}.
		The stars observed by the Bench spectrograph/WIYN are dominant in the lower-temperature regime.
		Because the WIYN telescope aperture is larger than that of the other telescopes, it enables us to observe fainter stars.
			
		The typical errors in $T_{\mathrm{eff}}$ are 128 K for KOOLS and KOOLS-IFU, and 85 K for the other instruments.
		We note that the spectra observed with KOOLS-IFU have low S/N (average S/N at 7400 $\mathrm{\AA} \sim 36$) and it results in larger errors for KOOLS-IFU. 
		In other studies, the $T_{\mathrm{eff}}$ of the M dwarfs have been determined with a precision of approximately 100 K or less.
		Mann et al. (2013b; 2015) determined the $T_{\mathrm{eff}}$ of M dwarfs from optical and near-infrared spectra with a precision of approximately 60 K.
		Using a wider wavelength coverage among our adopted ranges, we can determine the $T_{\mathrm{eff}}$ with a similar precision as that used in other studies.
		We note that our linear correction that is applied to the temperature estimations for the wavelengths of 5000--8000 \AA\ has larger uncertainties at the $T_{\mathrm{eff}}$ range $<$ 2600 K, suggesting that one needs to notice the derived temperatures of those (see also Section \ref{sec:analysis_teff_linearcorrection}).
		
		We also obtained $T_{\mathrm{eff}}$ of our M dwarfs by SED fitting, and an error in $T_{\mathrm{eff}}$ obtained from SED fitting is $\sim$ 40 K on average.
		We note that the errors should be underestimated, because we did not reflect the various uncertainties of model spectra into our analysis.
		The $T_{\mathrm{eff}}$ measured from SED are also summarized in Figure \ref{fig_result_histogram} and table \ref{tbl_result}.
		In table \ref{tbl_result}, there are some stars whose $T_{\mathrm{eff}}$ are not calculated from SED, because we calculated the $T_{\mathrm{eff}}$ for the stars whose photometric magnitudes and parallaxes are available.

	 \subsection{Radii \label{sec:result_radius}}
		We also obtained the stellar radii of our samples from their SED calculations.
		Stars with approximately 0.2 $R_\odot$ are the most common in our sample (see Figure \ref{fig_result_histogram}).
		The relative average error in radii is 3\%.
		As well as our $T_{\mathrm{eff}}$ estimation, we do not consider the incompleteness of the model spectrum in our analysis.
		Therefore, the errors on radii should be also underestimated.
		 
		There are some stars whose radii are larger than 0.6 $R_\odot$.
		LSPM J0919+6203 has a radius of $R =0.88 R_\odot$.
		However, its spectral type is estimated as M1 V and the parallax is 18.8578 mas, which is relatively small in our samples (the median of the parallaxes is 44.4032 mas).
		LSPM J1334+2011 also has a large radius (0.72 $R_\odot$) and its parallax is 18.2323 mas.
		Therefore, these stars may have large radii.

		LSPM J1914+1919 also has a large radius (0.61 $R_\odot$).
		Our $T_{\mathrm{eff}}$ estimation from its SED suggested $T_{\mathrm{eff}}=$ 3367 K, which is applied to the empirical relation in \citet{Mann2015}.
		As a result, the radius of LSPM J1914+1919 was inferred to be 0.32 $R_\odot$; this implies that we overestimated the radius of this object.
		LSPM J1914+1919 was reported as a multiple-star system in WDS (Mason et al. 2001, 2014), likely leading to an overestimate in the radius estimation.
		 
		As well as our $T_{\mathrm{eff}}$ estimation, we do not consider the incompleteness of the model spectrum in our analysis.
		Therefore, the errors on radii should be also underestimated.
		In addition, our result may be biased by unresolved binaries.

	\subsection{Spectral type}
		We show the histogram of the spectral types at the bottom panel in Figure \ref{fig_result_histogram}.
		Our results are also summarized in table \ref{tbl_result}.
		M4--M5 stars are dominant in our samples, and there are 17 late M dwarfs (later than M6) in our samples.
		The average error is 0.4 subtype, and it indicates that we can determine spectral types with high precisions.

\section{Discussion \label{sec:discussion}}
	
	\subsection{Temperature estimation for the KOOLS spectra}
\begin{figure}
			\centering
			\includegraphics[width=\columnwidth]{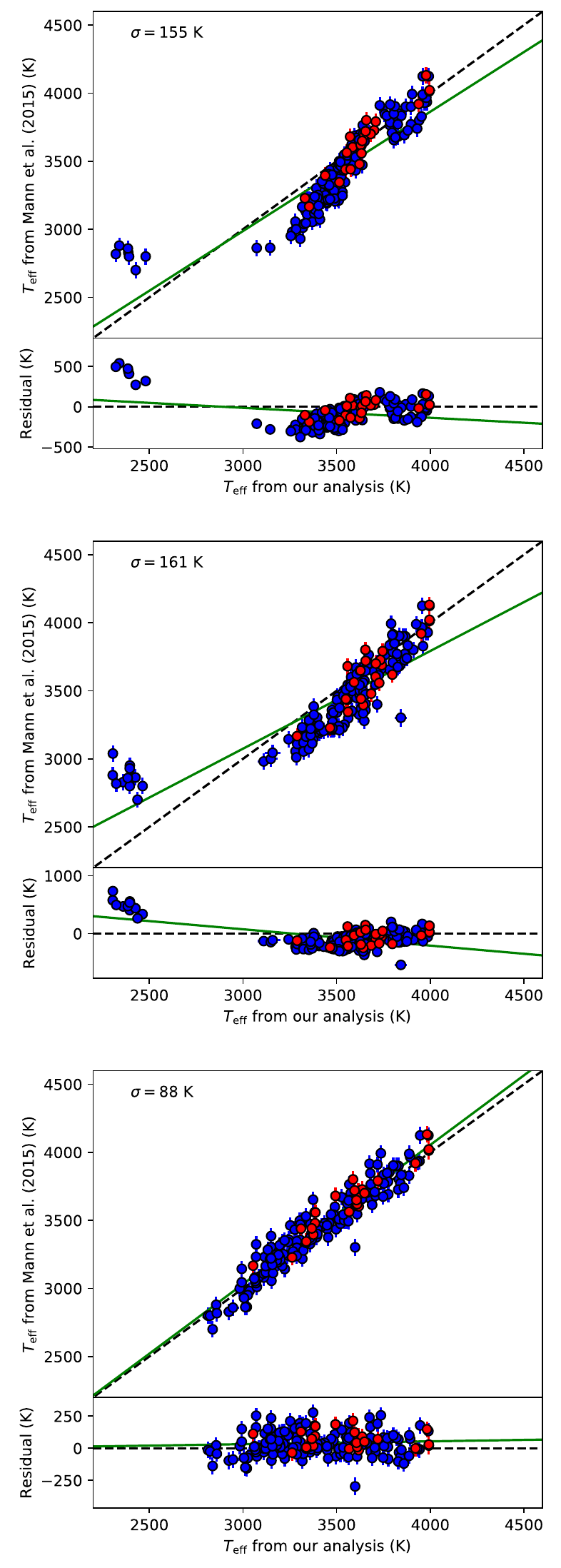}
			\caption{
				Same as Figure \ref{fig_teff_comparison_mann}, but temperatures from synthetic spectra using PHOENIX BT-Settl version \citep{Allard2011}.
				Top panel is TiO band1, middle is TiO band2, and bottom is VO band.
				\label{fig_teff_comparison_narrowband_BT}
			}
		\end{figure}

		In our analysis, we utilized the VO band to determine $T_{\mathrm{eff}}$ with the PHOENIX-ACES model grid \citep{Husser2013}.
		We also repeated our analysis for the spectra of \citet{Mann2015} with the PHOENIX BT-Settl version \citep{Allard2011}.
		We found that the resulted scatters of linear correction were 154 K for TiO band1, 161 K for TiO band2 and 88 K for VO band (see Figure \ref{fig_teff_comparison_narrowband_BT}).
		Again, the VO band gives the best-fitted results, though we employed the different models.
		We therefore concluded that TiO band1 should be useless for estimation of $T_{\mathrm{eff}}$ in spite of the differences of model spectra, with the caveat that our adopted two models are both the series based on the scheme of the PHOENIX group.
		This may be because the absorption of this band is shallow, and has been poorly constrained with respect to its line list as briefly described in section \ref{sec:analysis_teff_modelfit}.
	
	\subsection{Comparison between $T_{\mathrm{eff}}$ from optical spectra and from SED}
\begin{figure}
			\centering
			\includegraphics[width=\columnwidth]{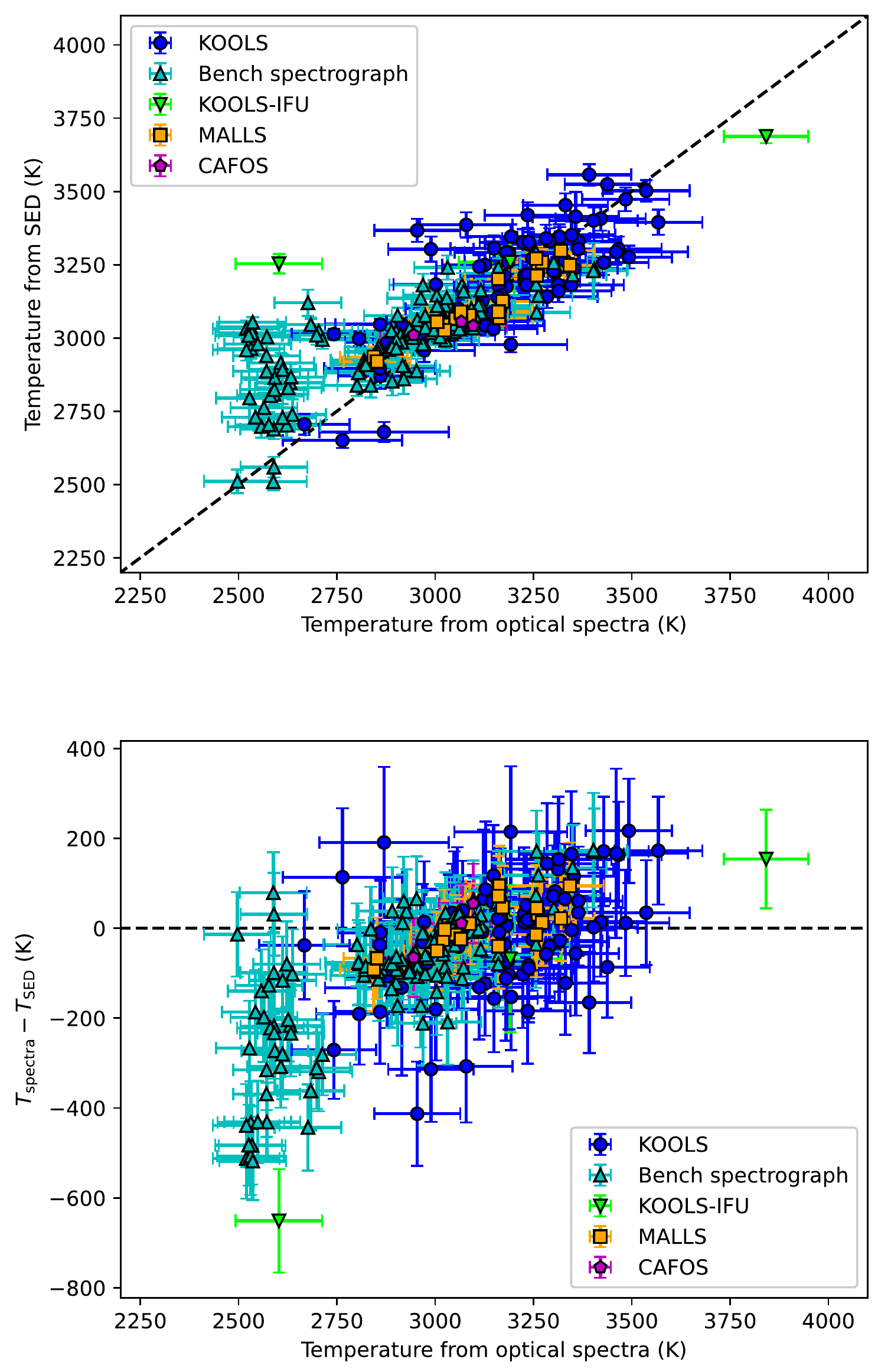}
			\caption{
				Comparison between $T_{\mathrm{eff}}$ from optical spectra and from SED.
				In top panel, vertical axis shows $T_{\mathrm{eff}}$ from SED and horizontal axis shows $T_{\mathrm{eff}}$ from optical spectra.
				Bottom panel shows the residual of the $T_{\mathrm{eff}}$.
				\label{fig_teff_comparison_modelfit_sed}
			}
		\end{figure}

		We compared the $T_{\mathrm{eff}}$ from optical spectra with those from SED in Figure \ref{fig_teff_comparison_modelfit_sed}.
		In this comparison, we exclude the stars with S/N at 7400 $\mathrm{\AA}<50$.
		The standard deviation of the difference between the $T_{\mathrm{eff}}$ derived by the two methods is 165 K.
		Most of our results follow the 1:1 relation.
		However, we found a trend towards higher $T_{\mathrm{eff}}$.
		In the lower $T_{\mathrm{eff}}$ region, $T_{\mathrm{eff}}$ from optical spectra tend to have smaller values than those from SED.
		Our linear correction for wavelength range of 5000--8000 $\mathrm{\AA}$ has larger deviation in lower $T_{\mathrm{eff}}$ region (see top panel in Figure \ref{fig_teff_comparison_mann}).
		This deviation could be attributed to the larger deviations of our derived $T_{\mathrm{eff}}$ from the ones in \cite{Mann2015} (see section \ref{sec:analysis_teff_linearcorrection}) or a general incompleteness of modeling such cool atmospheres.
		The removal of stars with $T_{\mathrm{eff}} <$ 2600 K yields a standard deviation of 130 K in the differences of the spectrum-based $T_{\mathrm{eff}}$ and the SED-based one.
		We cannot find clear reason for trend towards higher $T_{\mathrm{eff}}$ and outliers.
		Unlike methods which utilize high resolution spectroscopy, we do not fit the continuum level.
		Accuracy of flux calibration is vital for determination of $T_{\mathrm{eff}}$.
		Thus, there might be some incompleteness of flux calibration in our spectra.

	\subsection{Comparison with other studies}
		\subsubsection{Temperature}
\begin{figure}
				\centering
				\includegraphics[width=\columnwidth]{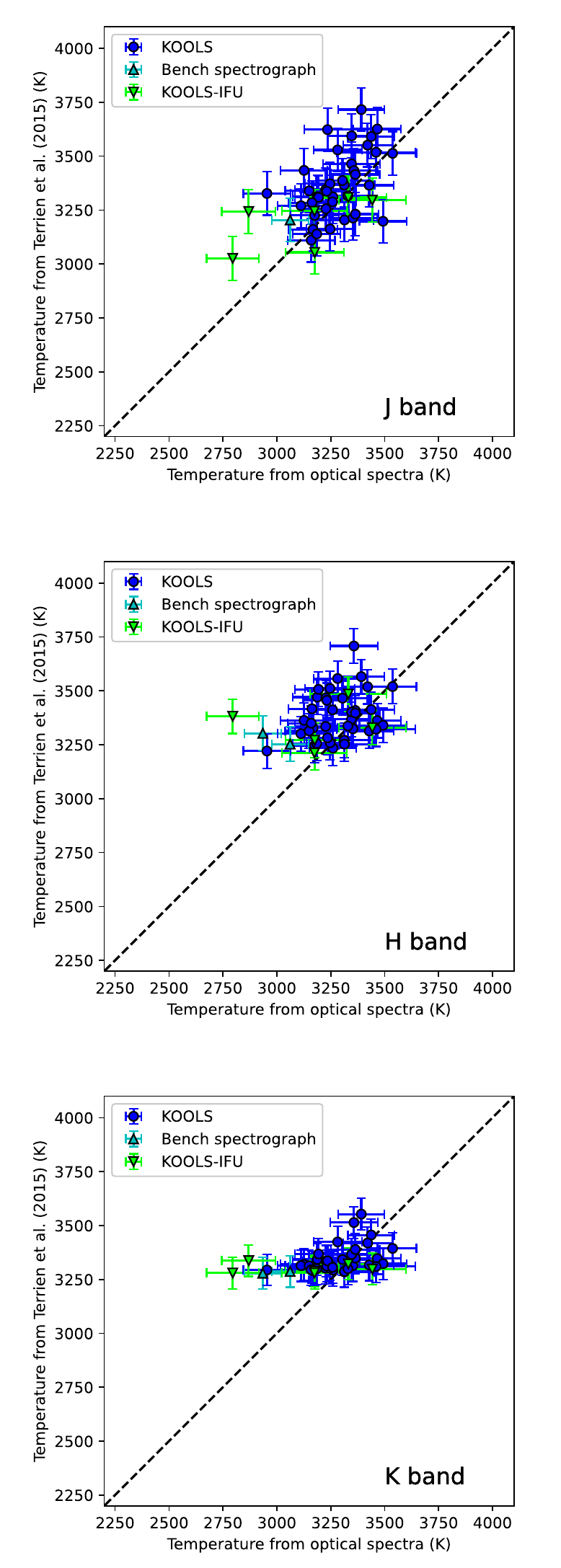}
				\caption{
					Comparison of our results with \citet{Terrien2015}.
					Top panel: comparison with $T_{\mathrm{eff}}$ from  $J$ band, middle panel: $H$ band, and bottom panel: $K$ band.
					In all panels, Vertical axis shows $T_{\mathrm{eff}}$ from \citet{Terrien2015} and horizontal axis shows $T_{\mathrm{eff}}$ from our analysis.
					\label{fig_teff_comparison_terrien}
				}
			\end{figure}

			We compare our result with \citet{Terrien2015} using the 62 common stars.
			They derived the $T_{\mathrm{eff}}$ of M dwarfs from the spectral features in the near-infrared using the methods of \citet{Mann2013b}.
			We show the comparison in Figure \ref{fig_teff_comparison_terrien}.
			In \citet{Mann2013b}, the relations are derived between the $T_{\mathrm{eff}}$ and molecular absorption features in the $JHK$ bands.
			The $T_{\mathrm{eff}}$ is expressed as a quadratic polynomial of the absorption lines in each band.
			However, their relations are based on the stars with 3200--5000 K, and therefore the $T_{\mathrm{eff}}$ below 3200 K may not be calculated.
			We conclude that our result is more accurate that and our methods are applicable to a wider $T_{\mathrm{eff}}$ range.

		\subsubsection{Radii}
\begin{figure}
				\centering
				\includegraphics[width=\columnwidth]{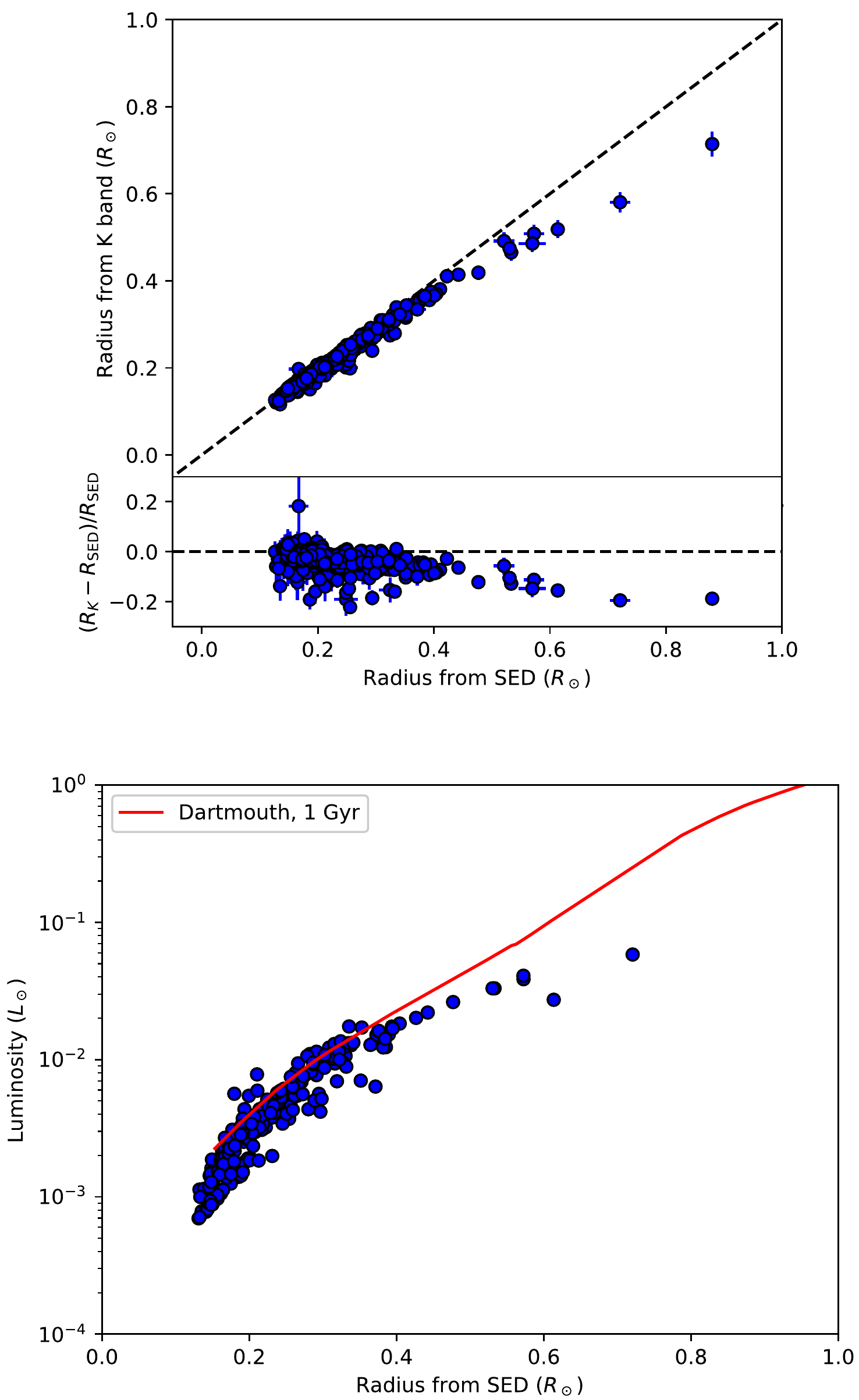}
				\caption{
					Comparisons with empirical radius (top) and Dartmouth Stellar Evolution Database (bottom).
					Empirical radius is calculated from mass-luminosity relation of \citet{Mann2019} and mass-radius relation of \citet{Rabus2019}.
					In the bottom panel, solid line represents 1 Gyr Dartmouth evolution model \citep{Dotter2008}.
					\label{fig_radius_comparison}
				}
			\end{figure}

			We compared our radii with the empirical radii.
			Gaia provides the accurate trigonometric parallaxes of our samples.
			Therefore, we could obtain the luminosities in the $K$ band by combining the Gaia parallaxes and 2MASS observations.
			\citet{Mann2019} derived an empirical relation between the mass and $K$ band absolute magnitude.
			\citet{Rabus2019} obtained an empirical mass-radius relation for low-mass stars.
			From these two empirical relations, the radii of our samples were calculated.
			The comparison is shown in Figure \ref{fig_radius_comparison} and our radii are on average 3\% larger than the empirical radii.
			We also found larger disagreements in larger radii ($>0.6R_\odot$).
			However, their radii are overestimated (see section \ref{sec:result_radius}).
			We also compared our results with the stellar evolution model.
			Figure \ref{fig_radius_comparison} shows the comparison of the radii and luminosities of our samples with the Dartmouth Stellar Evolution Database \citep{Dotter2008}.
			The $[\rm{Fe/H}]$ of the evolution model is fixed at the solar value, and we set the stellar age to be 1 Gyr.
			We found that our radii are also approximately 9\% larger than those from the stellar evolution model.
			Although it is unclear and beyond the scope of this paper why the radii of our samples were overestimated relative to the predictions from stellar evolution models, the reason might be the absence of strong magnetic fields in the models.
			\citet{Mullan2001} proposed that an inhibition of convection in stars with strong magnetic activities makes their radii larger.
			Strong magnetic fields should be produced in rapidly-rotating stars, and close binary systems are tidally locked and rapidly rotating (see \cite{Chabrier2007}; \cite{Kesseli2018} and references therein).
			However, \citet{Kesseli2018} inferred the radii of rapidly-rotating mid-to-late M dwarfs with inflations compared with the model predictions but without significant inflations compared with slowly-rotating stars, concluding that their overestimated radii are not due to rapid rotation (and binarity).
			Our overestimation is about 9\%, which is close to that in Kesseli et al. 2018 (10--15\%).
			Assuming most of our samples to be slowly-rotating inactive stars, we also find that rapid rotation (i.e., strong magnetic field) should not be mainly responsible for their overestimated radii.
			This means that stellar evolutionary models can underpredict the radii of inactive M dwarfs, consistent with the finding of \citet{Mann2015} that radius underpredictions of evolution models do not correlate with activity indicators.

		\subsubsection{Spectral type}
\begin{figure}
				\centering
				\includegraphics[width=\columnwidth]{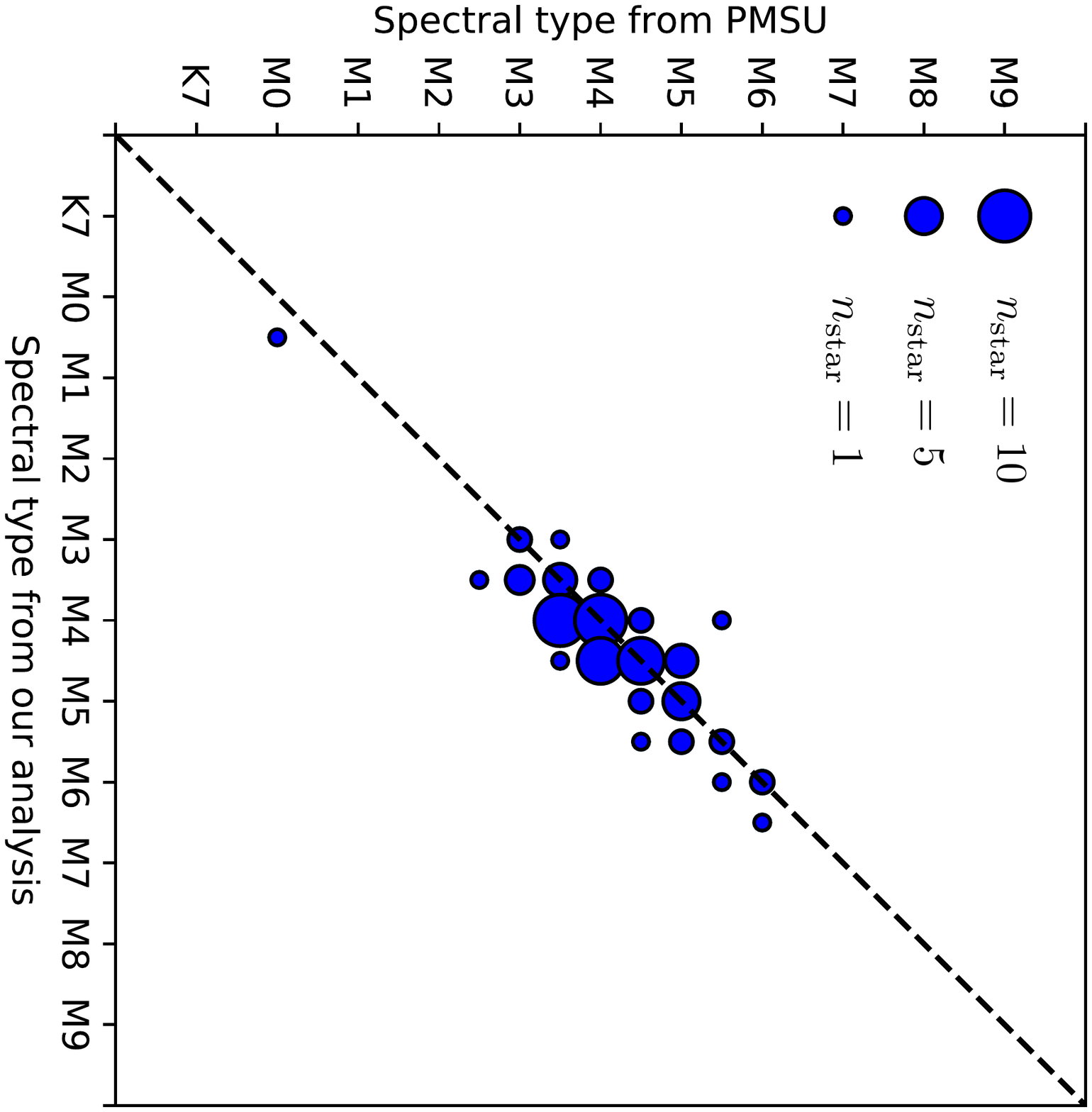}
				\caption{
					Comparison of our results with PMSU.
					The vertical axis shows spectral types from PMSU's and the horizontal axis shows spectral types from our analysis.
					The larger a circle, the greater the number of stars on a data point.
					\label{fig_spt_comparison}
				}
			\end{figure}

			We compared our results with PMSU using 79 common stars.
			In Figure \ref{fig_spt_comparison}, we show that our results are in agreement with PMSU.
			The standard deviation of the difference between the spectral types derived by us and by PMSU is 0.4 subtype.
			The average error in our result is 0.4 subtype, and we can derive the spectral types with the same precision as in previous research (e.g., \cite{AlonsoF2015}).
			Thus, we conclude that our results are reliable.

	\subsection{Effect of metallicity}
\begin{figure}
			\centering
			\includegraphics[width=\columnwidth]{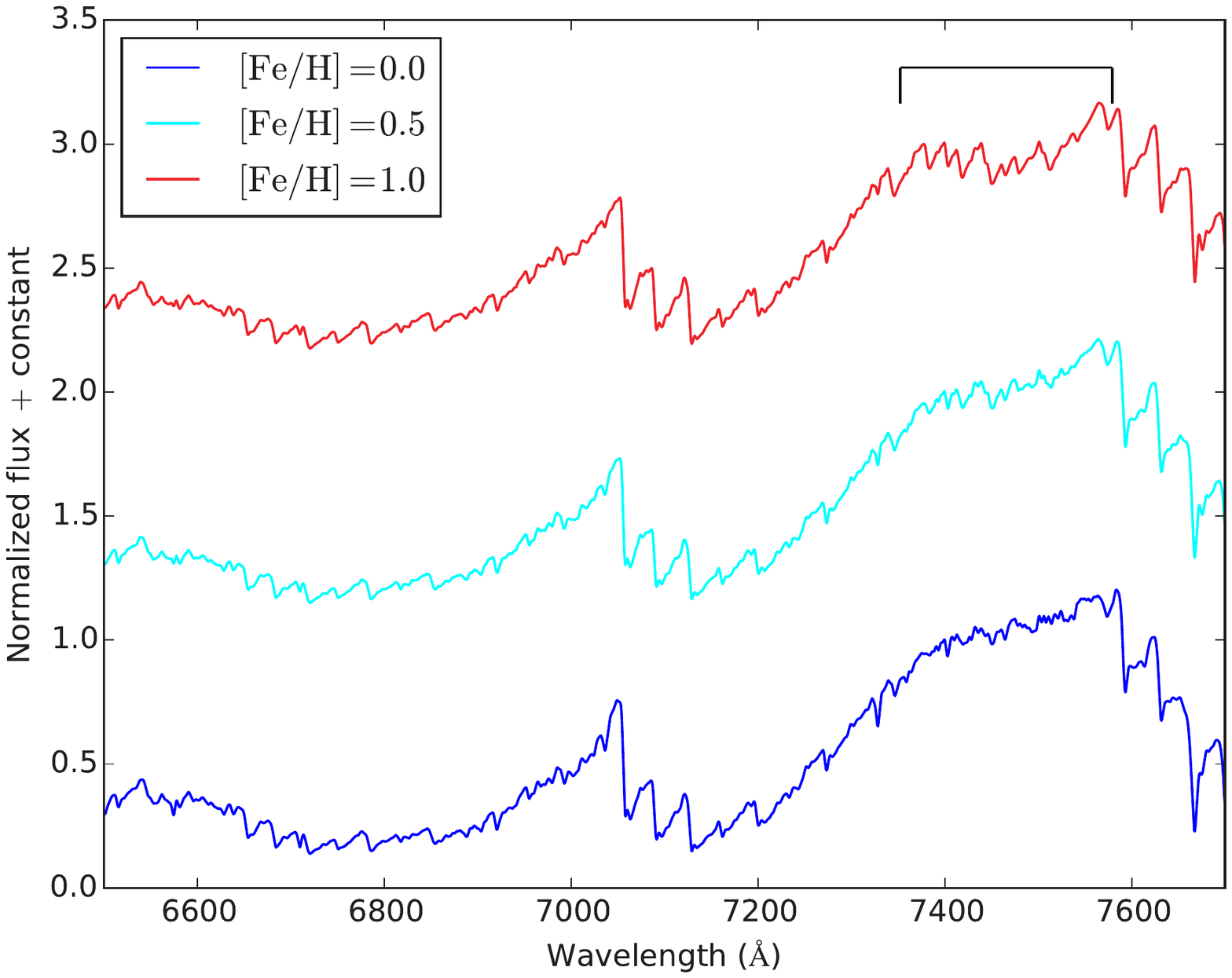}
			\caption{
				Synthetic spectra of [Fe/H] $=$ 0.0, 0.5, and $-$1.0 taken from \citet{Husser2013} that we used in our $T_{\mathrm{eff}}$ estimations.
				The $T_{\mathrm{eff}}$ is fixed as $T=3000$ K in this plot.
				The VO band is a region in 7320--7570 \AA.
				\label{fig_model_fehdependence}
			}
		\end{figure}

		Metallicities have been estimated from optical spectra for early M dwarfs, and from infrared spectra for mid-to-late M dwarfs (e.g., Mann et al. 2013a; 2014).
		We used the VO band to determine the $T_{\mathrm{eff}}$ for the data of KOOLS and KOOLS-IFU, but the VO band is also known to be sensitive to metallicity \citep{Mann2013a}.
		In Figure \ref{fig_model_fehdependence}, we showed the PHOENIX model spectra that were used in section \ref{sec:analysis_teff_modelfit}.
		As can be seen in Figure \ref{fig_model_fehdependence}, VO absorption appears as the metallicity increases.
		Therefore, our $T_{\mathrm{eff}}$ may be affected by metallicity.
		We recalculated the $T_{\mathrm{eff}}$ from optical spectra, fixing $[\rm{Fe/H}]$ as 0.0.
		For the data of KOOLS and KOOLS-IFU, the $T_{\mathrm{eff}}$ of 37\% of stars changed by $\leq100$ K, and of 13\% of stars changed by $\geq300$ K.
		The mean change in the data of KOOLS and KOOLS-IFU is 179 K.
		For the stars observed with the other instruments, the $T_{\mathrm{eff}}$ of their 76\% changed by $\leq100$ K, and the mean change is 89 K.
		The $T_{\mathrm{eff}}$ can be changed on the order of 100 K, and the influence of metallicity can be mitigated if a wider wavelength range is available.
		As demonstrated in e.g., Mann et al. (2013a; 2014), molecular absorption features in near-infrared spectra should be more adequate for an M dwarf with respect to its metallicity estimation.
		Once such a method restricts the metallicity of our sample, its stellar properties can be improved by combining the metallicity restriction and our observed spectrum.

\section{Conclusion and summary \label{sec:conclusion}}
	We determined the $T_{\mathrm{eff}}$, radii, and spectral types of 338 M dwarfs.
	The $T_{\mathrm{eff}}$ were calculated by comparing their optical spectra with the synthetic spectra of \citet{Husser2013}.
	To estimate more precise $T_{\mathrm{eff}}$, we calibrated our method using the data of \citet{Mann2015}, which have well-determined $T_{\mathrm{eff}}$ values.
	Some of our spectra have a narrow-wavelength region (6400--7600 $\mathrm{\AA}$), for which we found that it is better to focus on one of three molecular absorption bands than using the entire wavelength regions; the VO band is the most sensitive to $T_{\mathrm{eff}}$.
	The typical errors were 128 K for KOOLS and KOOLS-IFU, and 85 K for the other instruments.
	Although the use of a wider wavelength coverage makes it more accurate and precise to estimate $T_{\mathrm{eff}}$ from optical spectra, we demonstrated that it is possible to reproduce $T_{\mathrm{eff}}$ by selecting suitable region even if only narrow wavelength coverage is available.
	This helps to infer temperature of M dwarfs when one can use only spectrographs with narrow wavelength coverages including VO absorption on 7320--7570 \AA.
	For the stars with $T_{\mathrm{eff}}<$ 2600 K, we note that our linear correction for 5000--8000 \AA\ has larger deviations and thus our method would fail to estimate correct $T_{\mathrm{eff}}$.
	We tested the effect of metallicity, and found that our $T_{\mathrm{eff}}$ estimations can change on the order of 100 K.
	The effect of metallicity is important if one hopes to estimate the $T_{\mathrm{eff}}$ of an M dwarf with the precision  better than 100 K, and it can be mitigated given a wider wavelength range.
	Near-infrared spectra can restrict the metallicity of M dwarfs (e.g., Mann et al. 2013a; 2014). 
	If near-infrared spectra can be combined with the optical spectra in this paper, more robust $T_{\mathrm{eff}}$ estimations are available.  
	
	We also calculated the $T_{\mathrm{eff}}$ from SED with the typical error of 40 K.
	In the comparison between the $T_{\mathrm{eff}}$ from optical spectra and those from SED, we found deviations at the higher $T_{\mathrm{eff}}$ region.
	Although we cannot find any clear reasons, the deviations arise at the spectral types of M4--M5.5; one needs to take care of the possibility that our $T_{\mathrm{eff}}$ estimations from optical spectra may be overestimated at these spectral types.
	We also obtained the radii of our samples by SED calculation.
	Stars with approximately 0.2 $R_\odot$ are the most common in our sample.
	The average error in the radii is 3\%.
	Comparing our results with the empirical radii and the stellar evolution model, we found that our radii are larger than the empirical and theoretical predictions.
	Finally, We determined the spectral types of our samples, which are consistent with the results of previous studies, and the errors of those spectral types are as small as in the research of \citet{AlonsoF2015}. 
	
	Any exoplanet surveys infer the properties of identified planets, such as  mass, $T_{\mathrm{eff}}$, and habitability. 
	Those characterizations depend on the properties of their parent stars.
	In addition, the formation and evolution history of planets are expected to be dependent on the properties of their host stars. 
	Accordingly, it is essential to obtain more precise and robust stellar parameters.
	As well, the stellar parameters have to be well constrained before the start of surveys, since target selection plays an important role in the strategy of an exoplanet survey.
	
	Our results can help to determine the stellar properties of M dwarfs that have been observed in exoplanet surveys, as well as their target selections. 
	Furthermore, we provide the method that enables us to infer the $T_{\mathrm{eff}}$ of M dwarfs over limited wavelengths of optical spectra, based on the samples well calibrated by \citet{Mann2015}.
	Our developed methods can be applied to the optical spectra of M dwarf that have similar wavelength ranges as ours, to infer their $T_{\mathrm{eff}}$ with the precision of $\sim$100 K. 
	We also note that $T_{\mathrm{eff}}$ can be determined with the precision of 128 K based on VO absorption around $\sim7500\mathrm{\AA}$ if the wavelength coverage is not same as the \citet{Mann2015}.
	Currently, M dwarfs are one of the primary targets not only in RV surveys as described in section \ref{sec:intro} but also in transit surveys (e.g., \cite{Delrez2018}). 
	This work is wished to help those surveys efficiently perform target selection and well constrain the parameters of their discovered planets.

\begin{ack}
	We would like to thank Andrew Mann for giving their spectra and Hiroki Ishikawa for helping us with observation.
	We also thank Diane Harmer, Susan Ridgway and telescope operator for helping us with observation at WIYN.
	This work was supported by JSPS KAKENHI Grant Numbers 16K17660, 19K14783 and 8H05442A.
	This research has made use of the SIMBAD database, operated at CDS, Strasbourg, France.
	This research has made use of the VizieR catalogue access tool, CDS, Strasbourg, France (DOI : 10.26093/cds/vizier). The original description of the VizieR service was published in A\&AS 143, 23.
	This research has made use of the Washington Double Star Catalog maintained at the U.S. Naval Observatory
	This research is based on observations at Okayama Astrophysical Observatory (OAO), which is operated by National Astronomical Observatory of Japan, Nishi-Harima Astronomical Observatory, which is operated by Center for Astronomy of University of Hyogo, Centro Astron\'{o}mico Hispano Alem\'{a}n (CAHA) at Calar Alto, which is operated jointly by the Max-Planck Institut f\"{u}r Astronomie and the Instituto de Astrof\'{i}sica de Andaluc\'{i}a, and Apache Point Observatory 3.5-meter telescope, which is owned and operated by the Astrophysical Research Consortium.
	Data presented herein were obtained at the WIYN Observatory from telescope time allocated to NN-EXPLORE through the scientific partnership of the National Aeronautics and Space Administration, the National Science Foundation, and NSF’s NOIRLab.
	The WIYN observing nights were obtained through the NOAO proposal with IDs of 16B-0275 and 17A-0380 (PI: Masashi Omiya).
	NOAO is operated by the Association of Universities for Research in Astronomy (AURA) under a cooperative agreement with the National Science Foundation.
	GradPak was made possible by the U.S. National Science Foundation (NSF)  grants ATI-0804576, AST-1009471, and AST-1517006, as well as gift funds from the UW-Madison Department of Astronomy.
	The APO observations were in part performed with the observing time that we pursed with the financial support from Astrobiology Center, NINS.
	We appreciate our anonymous referee whose comments and suggestions significantly improved the quality of this paper.
\end{ack}

\section*{Supporting information}
	A complete listing of tables \ref{tbl_sptstandards}--\ref{tbl_stellarparameters2} is available in the online version of this article.
	A portions of tables \ref{tbl_sptstandards}--\ref{tbl_stellarparameters2} are shown in this article for guidance regarding its form and content.

\appendix

\section{Flux normalization for temperature estimation \label{appedix_linearcorrection}}
\begin{figure*}[ht]
		\centering
		\includegraphics[width=16cm]{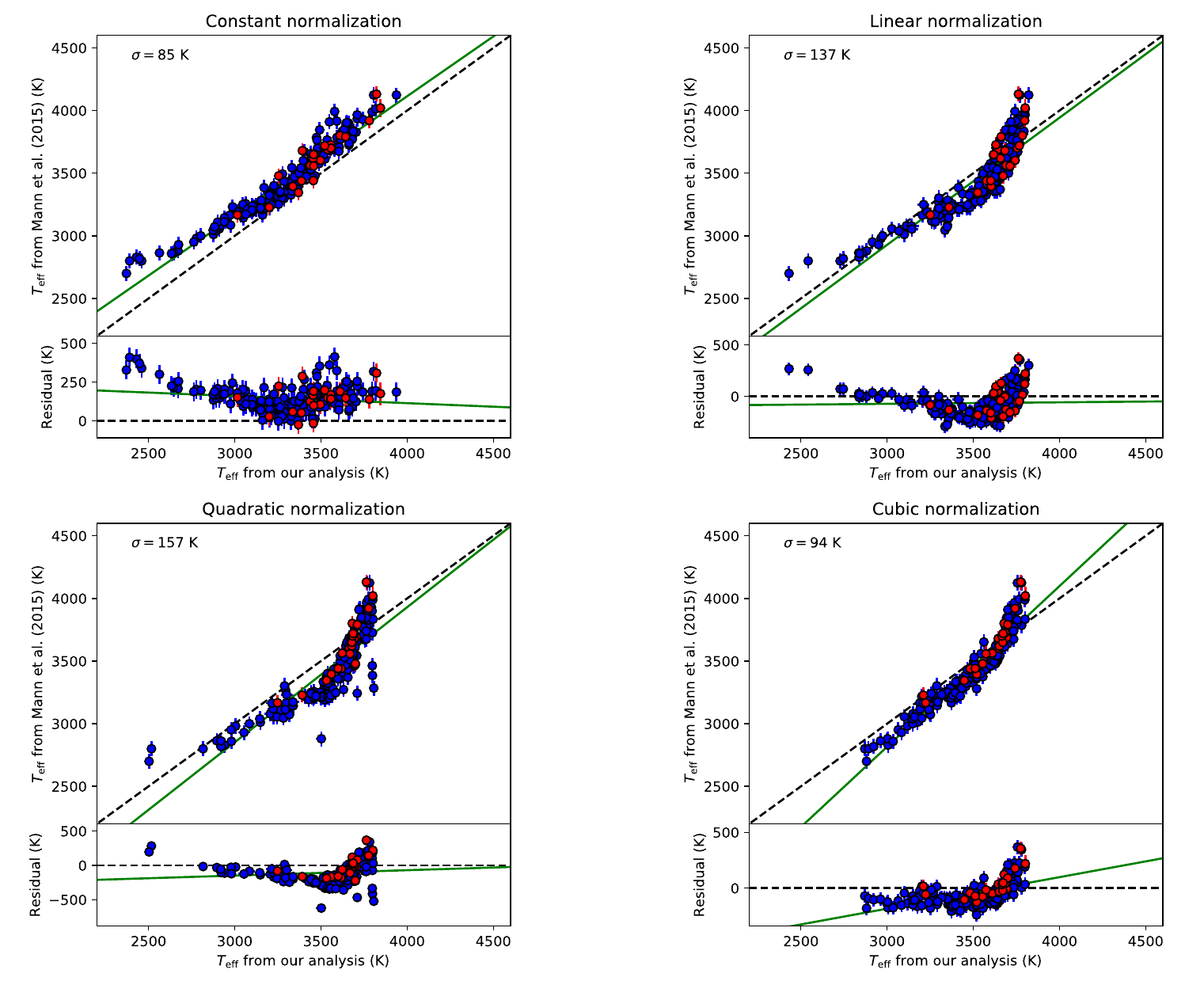}
		\caption{
			Same as Figure \ref{fig_teff_comparison_mann}, but only the wavelength range of 5000--8000 \AA\ is compared and the four types of normalizations are adopted in the optimizations of Equation (\ref{eq_chisq_spec}).
			Top left panel is the result of constant, top right is that of linear, bottom left is that of quadratic and bottom right is that of cubic normalization.
			Green lines in the figure represent linear correction functions (see section \ref{sec:analysis_teff_linearcorrection}) and $\sigma$ is the standard deviation of residual of the best-fit linear function.
			\label{fig_normalization}
			}
	\end{figure*}

	In the equation \ref{eq_chisq_spec}, we considered the constant, linear, quadratic, or cubic normalization.
	We adopted the constant normalization since it provides the smallest residuals (a standard deviation of 85 K) in the fit of the linear correction equation \ref{eq_linearcorrection} to the relations of our obtained T$_{\mathrm{eff}}$ and the T$_{\mathrm{eff}}$ from \cite{Mann2015} (see Figure \ref{fig_normalization}).
	In other words, constant normalization provides the most highest accuracy in the reproductions of well-calibrated $T_{\mathrm{eff}}$ estimations of \citet{Mann2015}.

\section{Spectral standard stars}
	Spectral standard stars that we used are listed\footnote{The table \ref{tbl_sptstandards} is available on the online edition as supporting information.} in table \ref{tbl_sptstandards}.
	\begin{longtable}{*{3}{c}}
		\caption{Spectral-type standards used in this study. \label{tbl_sptstandards}}
		\hline
		Spectral type & Object & Reference  \\
		\hline
		\endfirsthead
		\hline
		Spectral type & Object & Reference  \\
		\hline
		\endhead
		\hline
		\endfoot
		\hline
			K7 & BD+33 1505 & PMSU \\   
			K7 & HD 79210 & PMSU \\   
			K7 & HD 79211 & PMSU\\    
			M0.0 & BD+45 2743 & PMSU\\    
			M0.0 & HD 209290 & PMSU\\    
			M0.5 & BD+44 2051 A & PMSU\\    
			M0.5 & LHS 1747 & PMSU\\    
			M1.0 & BD+36 2219 & PMSU\\    
			M1.0 & GX And & PMSU\\    
			M1.5 & G 242-048 & PMSU\\    
	\end{longtable}

\section{Our samples}
	Star names, observation date, exposure times, and spectrographs are listed\footnote{The table \ref{tbl_observationlog} is available on the online edition as supporting information.} in table \ref{tbl_observationlog}.
	\begin{longtable}{*{4}{c}}
		\caption{Details of our observation. \label{tbl_observationlog}}
		\hline
		Object & Date & Exposure time (s) & Instruments  \\ 
		\hline
		\endfirsthead
		\hline
		Object & Date & Exposure time (s) & Instruments  \\ 
		\hline
		\endhead
		\hline
		\endfoot
		\hline
			EPIC 248536375 & 2018-01-26 & 2000 & DIS/APO 3.5 m \\
			GJ 1002 & 2018-01-25 & 600 & DIS/APO 3.5 m \\
			GJ 2005 & 2018-01-25 & 300 & DIS/APO 3.5 m \\
			GJ 3707 & 2016-04-15 & 600 & KOOLS-IFU/OAO 188 cm \\
			GJ 592 & 2016-04-15 & 900 & KOOLS-IFU/OAO 188 cm \\
			LEHPM 2-174 & 2017-05-05 & 1800 & Bench spectrograph/WIYN 3.5 m \\
			LP 851-346 & 2017-05-05 & 1800 & Bench spectrograph/WIYN 3.5 m \\
			LSPM J0020+3305 & 2016-11-20 & 400 & Bench spectrograph/WIYN 3.5 m \\
			LSPM J0027+4941 & 2014-09-22 & 600 & KOOLS/OAO 188 cm \\
			LSPM J0030+7742 & 2016-11-19 & 1140 & Bench spectrograph/WIYN 3.5 m \\
	\end{longtable}

\section{Results and stellar parameters}
	We summarize our results and stellar parameters\footnote{The table \ref{tbl_result}--\ref{tbl_stellarparameters2} is available on the online edition as supporting information.} in table \ref{tbl_result}--\ref{tbl_stellarparameters2}.
	$T_{\mathrm{optical}}$ is the temperature obtained from optical spectra.
	We note that the wavelength range of DIS is 6100 -- 7300 \AA, and we did not consider VO index for the spectra obtained by DIS.
	Therefore the spectral types of DIS data may be earlier than they actually are.
	\begin{longtable}{*{5}{c}}
		\caption{
			Stellar properties of our sample.
			The multipe-star systems listed in the WDS cataloggue (Mason et al. 2001, 2014) are marked with "*".
			\label{tbl_result}
		}
		\hline
		Object & Spectral type & $T_{\rm{optical}}$ (K) & $T_{\rm{SED}}$ (K) & Radius ($R_\odot$)  \\
		\hline
		\endfirsthead
		\hline
		Object & Spectral type & $T_{\rm{optical}}$ (K) & $T_{\rm{SED}}$ (K) & Radius ($R_\odot$)  \\
		\hline
		\endhead
		\hline
		\endfoot
		\hline
			EPIC 248536375 & $\mathrm{M2.5}\pm0.4$ & $\cdots$ & $\cdots$ & $\cdots$ \\
			GJ 1002 & $\mathrm{M5.5}\pm0.3$ & $\cdots$ & $2929\pm24$ & $0.1433\pm0.0048$ \\
			GJ 2005* & $\mathrm{M6}$ & $\cdots$ & $\cdots$ & $\cdots$ \\
			GJ 3707 & $\mathrm{M4}\pm0.4$ & $3620\pm119$ & $3237\pm37$ & $0.4107\pm0.0081$ \\
			GJ 592 & $\mathrm{M4}\pm0.3$ & $2603\pm110$ & $3254\pm33$ & $0.3459\pm0.0059$ \\
			LEHPM 2-174 & $\mathrm{M8.5}$ & $2588\pm85$ & $2509\pm29$ & $0.1863\pm0.0093$ \\
			LP 851-346 & $\mathrm{M8}$ & $2497\pm85$ & $2511\pm40$ & $0.1281\pm0.0087$ \\
			LSPM J0020+3305 & $\mathrm{M5.5}\pm0.3$ & $2631\pm85$ & $2847\pm36$ & $0.1464\pm0.0068$ \\
			LSPM J0027+4941 & $\mathrm{M4}\pm0.2$ & $3346\pm133$ & $3181\pm40$ & $0.3043\pm0.0082$ \\
			LSPM J0030+7742 & $\mathrm{M6}\pm0.4$ & $2558\pm85$ & $2698\pm37$ & $0.1616\pm0.009$ \\
			LSPM J0032+5429 & $\mathrm{M4}\pm0.2$ & $3313\pm141$ & $3182\pm41$ & $0.3174\pm0.0088$ \\
			LSPM J0035+0233* & $\mathrm{M5.5}\pm0.3$ & $\cdots$ & $2886\pm40$ & $0.2782\pm0.0153$ \\
			LSPM J0040+3122 & $\mathrm{M4}\pm0.3$ & $3305\pm123$ & $3223\pm44$ & $0.3383\pm0.0099$ \\
			LSPM J0044+0907 & $\mathrm{M4}\pm0.2$ & $3314\pm134$ & $3160\pm38$ & $0.4042\pm0.0103$ \\
			LSPM J0046+4851 & $\mathrm{M4.5}\pm0.4$ & $\cdots$ & $3029\pm25$ & $0.2817\pm0.0096$ \\
			LSPM J0049+6205 & $\mathrm{M5.5}\pm0.5$ & $2520\pm85$ & $3032\pm27$ & $0.1714\pm0.0065$ \\
			LSPM J0051+4531 & $\mathrm{M4.5}\pm0.3$ & $3405\pm85$ & $3235\pm46$ & $0.303\pm0.0097$ \\
			LSPM J0055+1439 & $\mathrm{M4}\pm0.4$ & $\cdots$ & $3193\pm40$ & $0.3287\pm0.0083$ \\
			LSPM J0100+6656 & $\mathrm{M4}\pm0.4$ & $3188\pm114$ & $3291\pm43$ & $0.3664\pm0.009$ \\
			LSPM J0101+3832 & $\mathrm{M5}\pm0.3$ & $3320\pm85$ & $3298\pm41$ & $0.3726\pm0.0085$ \\
			LSPM J0102+1009W* & $\mathrm{M1}\pm0.3$ & $\cdots$ & $3593\pm55$ & $0.5209\pm0.0184$ \\
	\end{longtable}
	
	\clearpage
	\clearpage
	\begin{table*}[h]
		\tabcolsep 1pt
		\scriptsize 
		\begin{center}
			\rotatebox{90}{
				\begin{minipage}{1.0\vsize}
					\caption{Stellar Parameters from other literatures.\footnotemark[$*$]   \label{tbl_stellarparameters1}}
					\begin{center}
						\begin{tabular}{ccc ccc c}
							\hline
							Object & Parallax (mas) & $V$ (mag) & $G$ (mag) & $G_{\mathrm{BP}}$ (mag) & $G_{\mathrm{RP}}$ (mag) & $J$ (mag) \\
							\hline 
								EPIC 248536375  & $\cdots$ & $\cdots$ & $\cdots$ & $\cdots$ & $\cdots$ & $\cdots$ \\
								GJ 1002  & $206.213\pm0.128$ & $13.837\pm0.030$ & $11.7804\pm0.0008$ & $14.0961\pm0.0024$ & $10.4046\pm0.0023$ & $8.323\pm0.019$ \\
								GJ 2005  & $\cdots$ & $15.301\pm0.030$ & $13.0942\pm0.0038$ & $15.5725\pm0.0063$ & $11.6056\pm0.0137$ & $9.254\pm0.034$ \\
								GJ 3707  & $81.552\pm0.104$ & $12.083\pm0.040$ & $10.7352\pm0.0010$ & $12.3571\pm0.0020$ & $9.5251\pm0.0010$ & $7.768\pm0.029$ \\
								GJ 592  & $71.655\pm0.080$ & $12.726\pm0.050$ & $11.3527\pm0.0017$ & $12.9767\pm0.0034$ & $10.1587\pm0.0017$ & $8.432\pm0.021$ \\
								LEHPM 2-174  & $56.081\pm0.252$ & $\cdots$ & $15.3726\pm0.0010$ & $18.5729\pm0.0250$ & $13.8258\pm0.0097$ & $11.160\pm0.023$ \\
								LP 851-346  & $91.595\pm0.154$ & $\cdots$ & $15.1217\pm0.0009$ & $18.4005\pm0.0222$ & $13.5791\pm0.0058$ & $10.930\pm0.023$ \\
								LSPM J0020+3305  & $81.631\pm0.117$ & $15.900\pm0.500$ & $13.9610\pm0.0005$ & $16.4199\pm0.0043$ & $12.5380\pm0.0030$ & $10.284\pm0.021$ \\
								LSPM J0027+4941  & $44.804\pm0.065$ & $14.192\pm0.050$ & $12.7892\pm0.0003$ & $14.4785\pm0.0022$ & $11.5369\pm0.0012$ & $9.733\pm0.021$ \\
								LSPM J0030+7742  & $76.000\pm4.000$ & $16.630\pm0.500$ & $14.3528\pm0.0013$ & $17.1088\pm0.0064$ & $12.8778\pm0.0027$ & $10.458\pm0.023$ \\
								LSPM J0032+5429  & $50.070\pm0.062$ & $13.859\pm0.030$ & $12.4587\pm0.0003$ & $14.1399\pm0.0019$ & $11.2063\pm0.0010$ & $9.387\pm0.022$ \\
								LSPM J0035+0233  & $38.127\pm0.373$ & $16.360\pm0.500$ & $14.2090\pm0.0035$ & $16.4858\pm0.0056$ & $12.7269\pm0.0036$ & $10.517\pm0.023$ \\
								LSPM J0040+3122  & $44.403\pm0.070$ & $13.798\pm0.040$ & $12.4890\pm0.0005$ & $14.0960\pm0.0030$ & $11.2642\pm0.0012$ & $9.491\pm0.022$ \\
								LSPM J0044+0907  & $37.572\pm0.345$ & $14.080\pm0.030$ & $12.6048\pm0.0004$ & $14.3415\pm0.0024$ & $11.3358\pm0.0016$ & $9.501\pm0.026$ \\
								LSPM J0046+4851  & $25.184\pm0.352$ & $16.590\pm0.500$ & $14.5870\pm0.0007$ & $16.5068\pm0.0080$ & $13.2723\pm0.0023$ & $11.339\pm0.021$ \\
								LSPM J0049+6205  & $37.893\pm0.059$ & $16.960\pm0.500$ & $14.7747\pm0.0005$ & $16.7483\pm0.0054$ & $13.4538\pm0.0020$ & $11.474\pm0.021$ \\
								LSPM J0051+4531  & $28.079\pm0.066$ & $15.007\pm0.040$ & $13.6986\pm0.0009$ & $15.2859\pm0.0062$ & $12.4728\pm0.0021$ & $10.715\pm0.022$ \\
								LSPM J0055+1439  & $24.769\pm0.080$ & $15.306\pm0.060$ & $13.8806\pm0.0007$ & $15.5607\pm0.0039$ & $12.6269\pm0.0028$ & $10.804\pm0.022$ \\
								LSPM J0100+6656  & $44.299\pm0.038$ & $13.425\pm0.020$ & $12.1603\pm0.0004$ & $13.6628\pm0.0022$ & $10.9736\pm0.0009$ & $9.408\pm0.029$ \\
								LSPM J0101+3832  & $26.570\pm0.076$ & $14.483\pm0.030$ & $13.2271\pm0.0013$ & $14.7573\pm0.0051$ & $12.0182\pm0.0036$ & $10.312\pm0.025$ \\
								LSPM J0102+1009W  & $18.858\pm0.045$ & $15.210\pm0.500$ & $12.6121\pm0.0006$ & $13.7365\pm0.0031$ & $11.5568\pm0.0016$ & $10.153\pm0.027$ \\ [2pt] 
							\hline\noalign{\vskip3pt} 
							\multicolumn{2}{@{}l@{}}{\hbox to0pt{\parbox{230mm}{\footnotesize
							\hangindent6pt\noindent
							\hbox to6pt{\footnotemark[$*$]\hss}\unskip%
							See section \ref{sec:analysis_sed} for the references of each photometric measurement. \\
							Parallaxes are from Gaia DR2  \citep{Gaia2018} or MEarth  \citep{Dittmann2014}. 
							}\hss}} 
						\end{tabular}
					\end{center}
				\end{minipage} 
			}
		\end{center}
	\end{table*}

	\clearpage 
	\begin{table*}[h]
		\tabcolsep 1pt
		\scriptsize
		\begin{center}
			\rotatebox{90}{
				\begin{minipage}{1.0\vsize}
					\caption{Stellar Parameters from other literatures.\footnotemark[$*$]   \label{tbl_stellarparameters2}}
					\begin{center}
						\begin{tabular}{ccc ccc c}
							\hline
							Object & $H$ (mag) & $K$ (mag) & $W1$ (mag) & $W2$ (mag) & $W3$ (mag) & $W4$ (mag) \\
							\hline 
								EPIC 248536375  & ... & ... & ... & ... & ... & ... \\
								GJ 1002  & $7.792\pm0.034$ & $7.439\pm0.021$ & $7.176\pm0.028$ & $6.993\pm0.020$ & $6.860\pm0.016$ & $6.766\pm0.080$ \\
								GJ 2005  & $8.547\pm0.036$ & $8.241\pm0.030$ & $7.836\pm0.023$ & $7.620\pm0.021$ & $7.353\pm0.017$ & $7.184\pm0.103$ \\
								GJ 3707  & $7.137\pm0.040$ & $6.863\pm0.024$ & $6.712\pm0.037$ & $6.555\pm0.020$ & $6.490\pm0.016$ & $6.384\pm0.054$ \\
								GJ 592  & $7.871\pm0.036$ & $7.572\pm0.023$ & $7.386\pm0.025$ & $7.252\pm0.021$ & $7.149\pm0.016$ & $7.066\pm0.098$ \\
								LEHPM 2-174  & $10.550\pm0.023$ & $10.128\pm0.021$ & $9.889\pm0.023$ & $9.678\pm0.021$ & $9.380\pm0.029$ & $8.857\pm0.332$ \\
								LP 851-346  & $10.295\pm0.023$ & $9.881\pm0.019$ & $9.646\pm0.021$ & $9.448\pm0.019$ & $9.235\pm0.030$ & ... \\
								LSPM J0020+3305  & $9.691\pm0.022$ & $9.330\pm0.018$ & $9.092\pm0.023$ & $8.901\pm0.020$ & $8.724\pm0.024$ & $8.227\pm0.189$ \\
								LSPM J0027+4941  & $9.160\pm0.021$ & $8.852\pm0.018$ & $8.665\pm0.022$ & $8.522\pm0.019$ & $8.376\pm0.019$ & $8.317\pm0.168$ \\
								LSPM J0030+7742  & $9.891\pm0.027$ & $9.562\pm0.021$ & $9.317\pm0.023$ & $9.115\pm0.019$ & $8.832\pm0.022$ & $9.036\pm0.378$ \\
								LSPM J0032+5429  & $8.827\pm0.016$ & $8.570\pm0.014$ & $8.406\pm0.022$ & $8.223\pm0.020$ & $8.118\pm0.018$ & $8.229\pm0.177$ \\
								LSPM J0035+0233  & $9.933\pm0.025$ & $9.543\pm0.021$ & $9.309\pm0.023$ & $9.110\pm0.021$ & $8.959\pm0.030$ & $8.845\pm0.448$ \\
								LSPM J0040+3122  & $8.864\pm0.030$ & $8.592\pm0.020$ & $8.419\pm0.024$ & $8.293\pm0.021$ & $8.158\pm0.019$ & $7.936\pm0.158$ \\
								LSPM J0044+0907  & $8.957\pm0.030$ & $8.621\pm0.020$ & $8.434\pm0.024$ & $8.273\pm0.019$ & $8.145\pm0.020$ & $8.105\pm0.228$ \\
								LSPM J0046+4851  & $10.733\pm0.019$ & $10.432\pm0.018$ & $10.234\pm0.022$ & $10.052\pm0.020$ & $9.858\pm0.031$ & ... \\
								LSPM J0049+6205  & $10.897\pm0.026$ & $10.620\pm0.020$ & ... & ... & ... & ... \\
								LSPM J0051+4531  & $10.165\pm0.023$ & $9.912\pm0.020$ & $9.763\pm0.024$ & $9.579\pm0.020$ & $9.448\pm0.027$ & ... \\
								LSPM J0055+1439  & $10.216\pm0.030$ & $9.934\pm0.019$ & $9.785\pm0.024$ & $9.631\pm0.020$ & $9.531\pm0.035$ & ... \\
								LSPM J0100+6656  & $8.726\pm0.031$ & $8.484\pm0.023$ & $8.262\pm0.022$ & $8.140\pm0.021$ & $8.054\pm0.020$ & $8.629\pm0.370$ \\
								LSPM J0101+3832  & $9.729\pm0.030$ & $9.475\pm0.018$ & $9.310\pm0.023$ & $9.172\pm0.020$ & $9.065\pm0.023$ & $8.872\pm0.318$ \\
								LSPM J0102+1009W  & $9.525\pm0.102$ & $9.312\pm0.027$ & $8.315\pm0.023$ & $8.277\pm0.020$ & $8.261\pm0.021$ & $8.031\pm0.222$ \\  [2pt] 
							\hline\noalign{\vskip3pt} 
							\multicolumn{2}{@{}l@{}}{\hbox to0pt{\parbox{230mm}{\footnotesize
							\hangindent6pt\noindent
							\hbox to6pt{\footnotemark[$*$]\hss}\unskip%
							See section \ref{sec:analysis_sed} for the references of each photometric measurement. \\
							Parallaxes are from Gaia DR2  \citep{Gaia2018} or MEarth  \citep{Dittmann2014}. 
							}\hss}} 
						\end{tabular}
					\end{center}
				\end{minipage}
			}
		\end{center}
	\end{table*}

\end{document}